\documentclass[fleqn,usenatbib]{mnras}

\DeclareRobustCommand{\VAN}[3]{#2}
\let\VANthebibliography\thebibliography
\def\thebibliography{\DeclareRobustCommand{\VAN}[3]{##3}\VANthebibliography}

\usepackage{graphicx}	
\usepackage{amsmath}	
\usepackage[export]{adjustbox}

\def\rco{r_{\rm co}}
\def\rci{r_{\rm ci}}
\def\rsh{r_{\rm sh}}
\def\vin{v_{r{\rm in}}}
\def\vou{v_{r{\rm ou}}}
\def\rou{r_{\rm ou}}
\def\theou{\Theta_{\rm ou}}
\def\lou{\lambda_{\rm ou}}
\def\thein{\Theta_{\rm in}}

\def\mbh{M_{\rm BH}}
\def\rin{r_{\rm in}}
\def\rg{r_{\rm g}}
\def\roi{\rho_{\rm ou}}
\def\robg{\rho_{\rm bg}}
\def\rms{r_{\rm ms}}
\def\lsim{\lower.5ex\hbox{$\; \buildrel < \over \sim \;$}}
\def\gsim{\lower.5ex\hbox{$\; \buildrel > \over \sim \;$}}
\def \simeq{\lower.3ex\hbox{$\; \buildrel \sim \over - \;$}}

    \title[Oscillating shocks in accretion disc]{Oscillating shocks in the transonic viscous, variable $\Gamma$ accretion flows around black holes}

\author[S. Debnath et al.]{
Sanjit Debnath,$^{1,2}$\thanks{E-mail: sdebnath@aries.res.in(SD)}
Indranil Chattopadhyay,$^{1}$\thanks{E-mail: indra@aries.res.in(IC)}
Raj Kishor Joshi$^{1,3}$\thanks{E-mail: raj@aries.res.in(RKJ)}
\\
$^{1}$Aryabhatta Research Institute of Observational Sciences (ARIES), Manora Peak, Nainital 263002, India\\
$^{2}$Department of Applied Physics, Mahatma Jyotiba Phule Rohilkhand University, Bareilly, Uttar Pradesh, 243006, India \\ 
$^{3}$Department of Physics, Deen Dayal Upadhyaya Gorakhpur University, Gorakhpur, Uttar Pradesh 273009, India\\
}

\date{Accepted XXX. Received YYY; in original form ZZZ}

\pubyear{2023}

\begin{document}
\label{firstpage}
\pagerange{\pageref{firstpage}--\pageref{lastpage}}
\maketitle
\begin{abstract}
We investigate the time evolution of the transonic-viscous accretion flow around a non-rotating black hole. 
The input parameters used for the simulation are obtained from semi-analytical solutions.
This code is based on the TVD routine and correctly handles the angular momentum transport due to viscosity. The thermodynamic properties of the flow are described by {a variable adiabatic index} equation of state.
We regenerate the inviscid and viscous steady-state solutions, including shocks, using the simulation code and compare them with the semi-analytical solutions.
{The angular momentum piles up across a shock due to shock-jump conditions and viscous transport of angular momentum. 
This will push the shock-front outward and can result in shock oscillation or a complete destabilization of shock.
We study} how shocks behave in the presence of viscosity. As the viscosity parameter ($\alpha$) crosses a critical value, the previously steady shock becomes time-dependent, eventually leading to oscillations. The value of this critical viscosity depends on the injection angular momentum ($\lou$) and the specific energy ($\epsilon$). We estimated the posteriori bremsstrahlung and synchrotron cooling, and the net radiative output also oscillates with the frequency of the shock. We also study the variation of frequency, amplitude, and mean position of oscillation with $\alpha$. Considering a black hole with a mass of $10M_{\odot}$, we observed that the power spectrum exhibits a prominent peak at the fundamental frequency of a few to about tens of Hz, accompanied by multiple harmonics. This characteristic is frequently observed in numerous accreting black hole candidates.
\end{abstract}




\section{Introduction}Theory of accretion onto the compact objects started its journey with the seminal papers of \citep{1939PCPS...35..592H} and \citep{1952MNRAS.112..195B}. However, accretion physics gained popularity after the discovery of quasars and  X-ray binaries. In 1964, \citeauthor{1964ApJ...140..796S} computed the luminosity from accreting matter onto black holes by using the Bondi accretion model, but the luminosity obtained was orders of magnitude lower. Bondi flow is the radial inflow model; therefore, the gravitational energy released increases the infall velocity, which makes the infall timescale of accreting matter much smaller than the radiation cooling timescale. Hence, the radiated luminosity is low. In order to increase the infall time scale compared to the radiative cooling timescale, matter must possess angular momentum. From the black hole inner boundary condition, the matter is most likely to possess sub-Keplerian angular momentum closer to the horizon but may possess higher angular momentum at a larger radius. Therefore, a general accretion disc requires viscosity to effectively remove angular momentum from the system. The foundational work on a viscous accretion disc was presented by \cite{1973A&A....24..337S}. This model assumed a Keplerian angular momentum distribution, along with a geometrically thin and optically thick accretion disc. This disc is known as the Keplerian disc (KD). The spectra computed from the KD by summing up the black body emission from each of the annuli (each with different temperatures) matched the thermal part of the spectrum. As a result, KD is widely used to explain the observed optical continuum spectra of luminous active galactic nuclei (AGN) or the soft X-ray continuum spectra of the X-ray binaries. However, the KD did not adequately consider the pressure and advection terms, and it failed to satisfy the inner boundary condition near a black hole, except for the ad-hoc termination of the disc at $r \leq \rms$ (where $\rms$ represents the marginally stable orbit or the Innermost Stable Circular Orbit, ISCO). Moreover, KD could not reproduce the non-thermal power-law part of the spectrum associated with objects harbouring black holes. Therefore, the search for a general accretion model that can explain the high-energy non-thermal radiation was investigated by various groups. The next important model is called the thick disc model, also sometimes called the `Polish doughnut' model, was developed by \cite{1976ApJ...207..962F,1980AcA....30..347P,1980A&A....88...23P,1980ApJ...242..772A} etc. In this model, the pressure gradient force along the radial as well as the transverse direction was considered, although the advection term was neglected. This accretion model was rotation-dominated and hot; therefore, it was puffed up in the form of a torus. Hence the name `thick' accretion disc. Thick discs were shown to be susceptible to Papaloizou-Pringle (PP) instability \citep{1984MNRAS.208..721P}. However, it was also shown that for various configurations, thick discs are PP stable. 
Simultaneous with the development of these models, there were many efforts to develop models with advection terms present in the equations of motion. In 1980 \citeauthor{1980ApJ...240..271L} studied transonic, rotating accretion solutions in Schwarzschild metric. This paper invoked a Bondi-type transonic solution but in the presence of rotation. They also showed that, unlike the Bondi flow, a rotating transonic solution may admit multiple sonic points. Subsequent studies \citep{1987PASJ...39..309F, 1989ApJ...347..365C} have further demonstrated that two standing shocks can exist between {the} two {physical, X-type} critical points. {The inner shock jumped onto an accelerating part of the subsonic branch while the outer shock jumped onto the decelerating part of the same subsonic branch}. It was shown that the outer shock position out of the two available is stable, using analytical methods \citep{1992MNRAS.259..259N} and also by using smooth particle hydrodynamic simulations
\citep{1993ApJ...417..671C}. {The instability of the inner shock was also studied
in details in few other papers \citep{1993PASJ...45..167N,1994MNRAS.270..871N,1994PASJ...46..257N,1996MNRAS.281..226N}.}{\cite{2019PASJ...71...38F,2019MNRAS.483.3839F} has recently investigated various types of shock in accretion disc. \cite{2021MNRAS.506.5698F} calculated shock conditions in the accretion disc by considering $\alpha p$ viscosity prescription}. \cite{1995MNRAS.272...80C} proposed a new form of viscous stress tensor that conserved the angular momentum across any shock jump. Based on this prescription, a number of studies on viscous flows were undertaken \citep{1996ApJ...464..664C,2004ChPhL..21.2551G,2004MNRAS.349..649C,2007NewA...12..454C,2008NewA...13..549D}. However, many papers both in the analytical regime and numerical simulations, were also written using traditional viscosity prescription, where the viscous tensor was assumed to be proportional to the shear tensor \citep{1998MNRAS.299..799L,1999ApJ...523..340L,2008ApJ...677L..93B,2009ApJ...702..649D,2011ApJ...728..142L,2013MNRAS.430..386K,2014MNRAS.442..251D,2016ApJ...831...33L}.
One of the popular versions of accretion flows in the advective regime is called ADAF or Advection Dominated Accretion Flow \citep{1977ApJ...214..840I, 1994ApJ...428L..13N}, which is a radiatively inefficient flow in which most of the dissipated gravitational energy is advected to the central black hole. The initial solution for ADAF was self-similar and completely subsonic. 
However, subsequent global solutions of ADAF, considering the influence of strong gravity, demonstrated that the flow might be self-similar far from the central object but has to be transonic near the horizon \citep{1997ApJ...476...61C}. 
\cite{1999ApJ...523..340L} actually made a thorough classification of viscous, transonic solutions in the advective regime and showed that  ADAF solutions (monotonic, viscous accretion solutions with single sonic point closer to the horizon) are a part of the general advective solutions.

One of the major propositions for the advective transonic disc is the identification of the post-shock region to be the major contributor for inverse-Comptonized hot photons \citep{1995ApJ...455..623C}, in other words, the illusive Compton cloud. Even more interesting was the fact that \cite{1996ApJ...457..805M} showed that the post-shock disc could eject a part of the
accreting matter as bipolar jets. Moreover, \cite{1996ApJ...470..460M} showed resonant oscillation of the post-shock region, which might explain the low-frequency quasi-periodic oscillation (QPO). The resonant oscillation was achieved by the enhanced radiative cooling rates in the post-shock region. On the other hand, a number of simulations showed that viscosity could also trigger shock oscillation \citep{1998MNRAS.299..799L,2014MNRAS.442..251D,2011ApJ...728..142L,2015MNRAS.448.3221G,2016ApJ...831...33L}. These simulations showed that beyond a critical viscosity, the post-shock disc may oscillate which would produce an oscillation with similar frequency of high energy photons emitted by the post-shock disc. Impressive as these results might be, all of these simulations were launched with super-sonic injection, and therefore, those were not proper outer boundary conditions. It was pointed out in steady-state analytical works that shock location may or may not shift outwards {with the increasing viscosity parameter}. It all depends if the average angular momentum of the post-shock disc (PSD) is higher or lower than the angular momentum at the outer boundary\citep{2013MNRAS.430..386K}. Moreover, these simulations used a fixed adiabatic index ($\Gamma$) equation of state (EoS) of the gas that constitutes the accretion disc. It may be noted that an accretion disc may extend from near the black hole horizon to a few thousand Schwarzschild radii, and the temperature of the flow over this entire spatial range may vary for two to three orders of magnitude.
Over such a range of temperature distribution, the fixed $\Gamma$ EoS is not tenable since, in principle, $\Gamma$ is a function of temperature. It was also shown that $\Gamma$ is a function of temperature as well as the composition of the gas \citep{2008AIPC.1053..353C,2009ApJ...694..492C}.   

 In the present paper, we study time dependant transonic, viscous accretion flow onto a non-rotating black hole, using the variable $\Gamma$ EoS proposed by \cite{2008AIPC.1053..353C, 2009ApJ...694..492C} and its acronym is CR EoS. The angular momentum transport will depend on the local temperature, viscosity parameter, and the Mach number. It may be noted that supersonic flow is less susceptible to viscous momentum transfer compared to a subsonic flow. Since a CR EoS computes the temperature correctly, therefore, studying the effect of viscosity on flows with CR EoS is very important. Further, if one wants to study only shocked solutions, then supersonic injection may make partial sense. However, if the interest encompasses all kinds of accretion solutions
 including Bondi type, shocked as well as ADAF type solution then one has to study solutions with sub-sonic injection.
 Even for shocked accretion flow, supersonic injections produce insignificant angular momentum transport in the pre-shock part of the accretion disc, while there is significant angular momentum transport in the post-shock disc, which would render an angular momentum imbalance in the computation setup of the disc. In the case of the subsonic outer boundary condition, the angular momentum transport in the pre-shock disc (precisely, in the subsonic region beyond the outer sonic point) is {more than the supersonic part} and therefore, the average angular momentum distribution on either side of the shock-front might be comparable and therefore may stabilise the shock oscillation. 
CR EoS has been implemented in simulation codes studying various problems \citep{2013ASInC...9...13C,2022MNRAS.509...85J,2022ApJ...933...75J,2023ApJ...948...13J}.
However, since angular momentum is so important for accretion discs, we have
{replaced the azimuthal velocity by the specific angular momentum as one of the primitive variables} and used the corresponding eigenstructure to write the TVD code. Using this new code, we would first like to test the code in the steady state inviscid and viscous rotating accretion flows by comparing the simulation results with the semi-analytical solutions. In the process, we would like to check all possible accretion solutions in the advective regime, with subsonic outer boundary conditions. Our main
intention is to study time-dependent shocked accretion solutions for a wide range of different conditions. 
We would like to see how viscosity may affect the accretion shock in the time-dependent regime. If the shock is oscillating, we would like to {quantify} the range of viscosity parameter {which would admit shock oscillation}. And if the shock oscillation due to viscosity can be associated with
QPOs.

We have organised our paper as follows. We first describe the time-dependent conserved form of the equations of motion. Then, we present the equation of state (EoS), steady state semi-analytical viscous accretion flows solution in subsections 2.1 and 2.2.
In section 2.3 we discuss our simulation code. In section 3 we present the verification of our code considering both the inviscid as well as viscous flows. We present our results from the simulations in section 4. Finally, we discuss our results in section 5 in terms of the observed properties of the accreting black holes.    

\begin{figure}
	\centering
	\includegraphics[width=3.5 in]{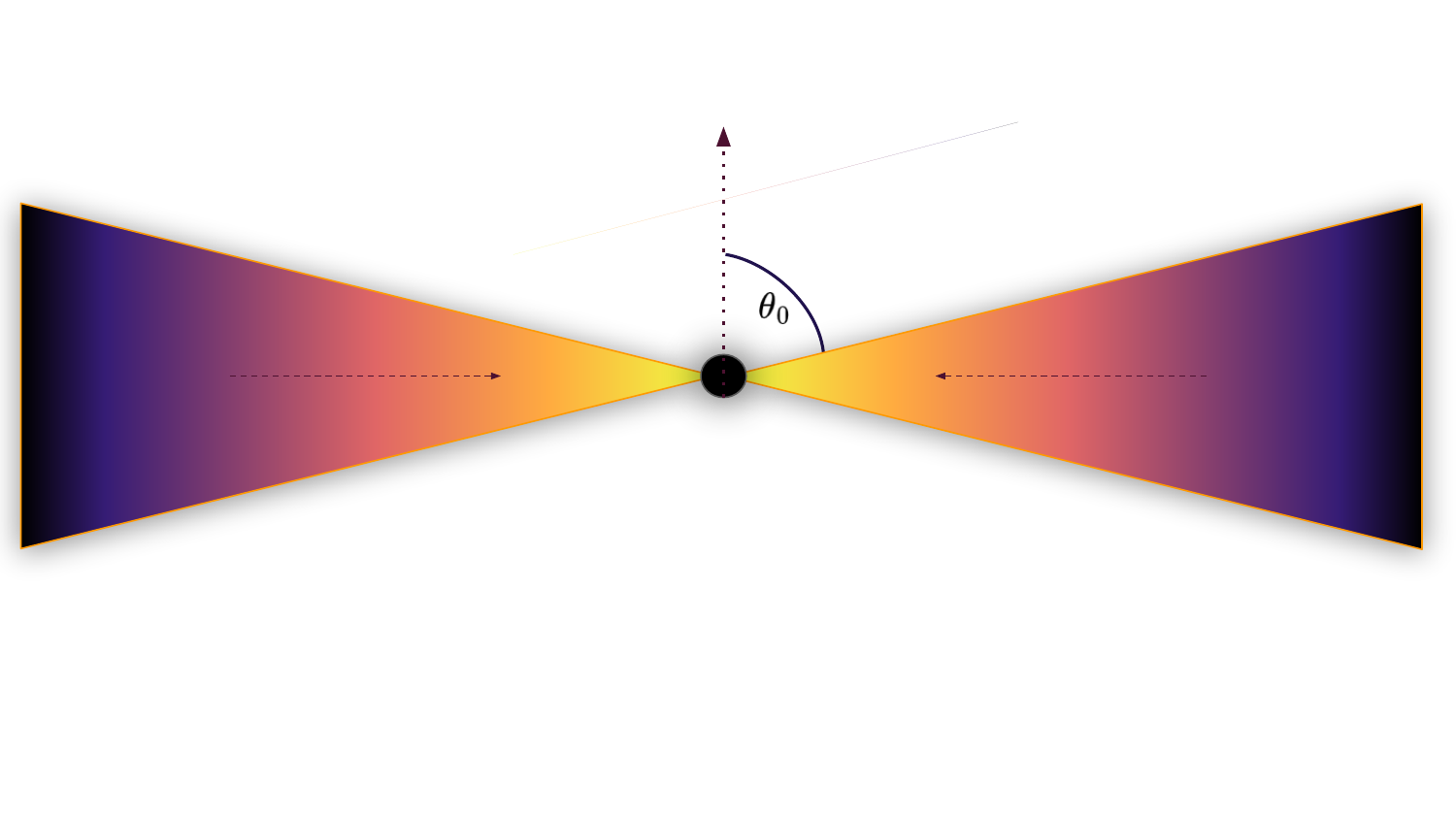}
        \caption{{Schematic representation of accretion disc geometry. Here $\theta_0$ is the co-latitude.}}
        \label{fig:geometry}
\end{figure}

\section{Assumptions and Equations }
This study focuses on examining the properties of viscous accretion flow around a non-rotating black hole, where the pseudo-Newtonian potential \citep{1980A&A....88...23P} is
 utilized to account for the strong gravitational field. {We consider that the flow is conical or wedge flow. The geometry of the accretion flows is shown in Fig.\ref{fig:geometry}, where $\theta_0$ is the co-latitude of the surface of the accretion disc. The central black spot represents the black hole (BH)}.
The fluid conserved variables or the state variables are represented as $\mathbf{q}$s, and the primitive variables are given as $\mathbf{w}$s,
\begin{equation}
\mathbf{q}= 
\begin{pmatrix}
\rho \\
 M_r \\
 M_{\theta} \\
  M_\lambda \\
E \\
\end{pmatrix}
=
\begin{pmatrix}
\rho \\
 \rho v_r \\
\rho v_{\theta} \\
 \rho \lambda \\
\rho v^2/2 +e \\
\end{pmatrix};
~~ \mathbf{w}= 
\begin{pmatrix}
\rho \\
 v_r \\
 v_{\theta} \\
 \lambda \\
 p \\
\end{pmatrix} 
\label{eq:state_primitv}
\end{equation}
Here, $\rho$ is the fluid rest mass density, {$M_r=\rho v_r$, $M_{\theta}=\rho v_\theta$ are the momentum density in the $r$ and $\theta$ directions. $M_{\lambda}=\rho \lambda$ is the angular momentum density, $E=\rho v^2/2+e$  represents the total energy density, which includes the internal energy and the kinetic energy. Here,} {$p$ is the pressure, $e$ is the energy density of the gas, $v_r$, $v_\theta$ and $v_{\phi}$ are velocity components along $r$, $\theta$ and $\phi$ while $\lambda=rv_{\phi}$ is the specific angular momentum}. {In addition $v^2=v_r^2+v_\theta^2+v_\phi^2$}. The equations of motion in the conserved form for 1D spherical {coordinate system} are given as
\begin{equation}
 \frac{\partial \mathbf{q}}{\partial t} + \frac{1}{r^2} \frac{\partial (r^2\mathbf{F^r})}{\partial r}  = \mathbf{S}
 \label{eq:conserve}
\end{equation}
Where, the fluxes corresponding to the $\mathbf{q}$s are given as
\begin{equation}
\mathbf{F^r}= 
\begin{pmatrix}
\rho v_r \\
v_r M_r+p \\
v_r M_{\theta} \\
v_r M_\lambda \\
(E+p)v_r\\
\end{pmatrix} , 
\end{equation}
and the source terms of the equations of motion are given as
\begin{equation}
\mathbf{S}= 
\begin{pmatrix}
0 \\
\frac{\rho v^2_{\phi}}{r} -\frac{G \mbh \rho}{(r-\rg)^2}+\frac{2p}{r} \\
0 \\
 S_\lambda \\
 -\frac{G \mbh \rho v_r}{(r-\rg)^2}+S_E \\
\end{pmatrix}
\label{eq:Smatrix}
\end{equation}
The set of equations represented by Eq. \eqref{eq:conserve} consists of the continuity, three momentum equations, and the energy equation. The second row of $\mathbf{S}$ contains the gravitational force density, which is presented as $G \mbh \rho/(r-\rg)^2$, obtained from the Pacz\'nsky-Wiita potential
\citep{1980A&A....88...23P}, which is given as $\Phi=-G \mbh/(r-\rg)$, where $G$ is the gravitational constant and $\mbh$ is the black hole mass and the Schwarzschild radius is $\rg=2 G \mbh/c^2$. Moreover, $\rho v^2_{\phi}/r$ is the force due to the rotation. Additionally, the angular momentum equation includes a source term arising from viscous dissipation, denoted as $S_\lambda$. The fifth row of $\mathbf{S}$ represents {the work done due to the gravity}, which is proportional to $G\mbh\rho v_r/(r-\rg)^2$, and {the energy source term is $S_E$}.
Here, $S_{\lambda}$ and $S_E$ are given as,
\begin{equation}
S_{\lambda}= {\frac{1}{r^2}\frac{\partial}{\partial r}\left(r^3 W_{r\phi}\right);} ~
~S_E= {(1-f_c)\frac{W^2_{r\phi}}{\eta_v}} 
\label{eq:sesl}
\end{equation}
{here, $f_c$ is cooling parameter, where $0\leq f_c<1$}. $W_{r\phi}$ is the $r - \phi$ component of viscous stress tensor. The local heat gained and lost by the flow are given by $Q^+=W^2_{r\phi}/\eta$ and $Q^-$. {Here $Q^-=f_c Q^+$. So $S_E=Q^+-Q^-$. We assume that the  $f_c=0$ in our study, until mentioned otherwise.} 
The viscous stress is given by,
\begin{equation}
W_{r\phi} = \eta_v r \frac{d\Omega}{dr}
\label{eq:stress}
\end{equation}
where, $\Omega$ is the angular velocity, $\eta_v = \rho \nu$ is the dynamic viscosity coefficient, {$\nu= (\alpha p)/(\rho \Omega_k)$} is the kinematic viscosity, $\alpha$ is the Shakura–Sunyaev viscosity parameter. $\Omega_k$ is the local Keplerian angular velocity and is given by,
\begin{equation}
\Omega^2_k = \frac{1}{r} \frac{d \Phi}{d r}
\label{eq:omegak}
\end{equation}
An additional closure relation for the thermodynamic variables $e, p, \rho$ is required to
solve the equations of motion (Eq. \ref{eq:conserve}) and is given by the CR EoS. 
\subsection{Equation of state} \cite{PhysRev.74.328} demonstrated that using a fixed $\Gamma$ EoS to describe the thermodynamics of flows with varying temperatures, which can differ by orders of magnitude, is unphysical. So it is not physical to describe the accretion around the black hole by a fixed $\Gamma$ equation of state (EoS) from infinity to horizon. To address this, we utilize an EoS for multispecies fluids with a variable adiabatic index proposed by \cite{2009ApJ...694..492C}, abbreviated as the CR EoS. Studies by \cite{2015MNRAS.453.2992V} have shown that the CR EoS is an excellent approximation of the exact EoS \citep{1939isss.book.....C}. 
{One may wonder why approximate EoS is at all used instead of the exact EoS! Approximate EoS were invoked in order to use it in a numerical simulation code. The exact EoS is a combination of modified Bessel's function. In simulation codes, one needs to re-compute the primitive variables from the state variables after those are updated by solving the hydrodynamic equations at every grid cell and at every time step. The exact EoS slows down simulation
codes significantly \citep{falle96, scheck02,pm07,mfs22}. It is for this reason an approximate and accurate modified EoS, which captures the physics correctly, has been pursued. CR EoS is extremely accurate and yet simple, to the point that the task of retrieving the primitive variables does not incur extra computational cost}.
Recently, \cite{2021MNRAS.502.5227J} presented an alternative form of the CR EoS, which we use in our analysis.
\begin{equation}
 e = \rho c^2 f
 \label{eq:eos}
\end{equation}
where, $f$ is given as 

\begin{equation}
 f = 1 + (2-\xi)\Theta\left[\frac{9\Theta+6/\tau}{6\Theta+8/\tau}\right]+\xi\Theta\left[\frac{9\Theta+6/\tau\eta}{6\Theta+8/\tau\eta}\right]
 \label{eq:f}
 \end{equation}
The fluid mass density, $\rho$ in the above equation can be expressed as $\rho=\sum n_i m_i=n_e m_e(2-\xi+\xi/\eta)$, where  $\eta= m_e /m_p$, $\xi =n_p/n_e$, and $n_p$, $n_e$, $m_p$, and $ m_e $ presented the proton number density, the electron number density, the proton rest mass, and electron rest mass, respectively. In this paper, we assume that the accretion flow consists solely of protons and electrons $(e^--p^+)$. As a result, the value of $\xi$ is 1. Additionally, $\Theta= p/\rho c^2$ serves as a measure of temperature and $\tau = 2-\xi + \xi/\eta$. The expression for specific enthalpy is also provided as,
\begin{equation} 
 h=(e+p)/\rho=(f+\Theta)c^2
 \label{eq:h}
\end{equation}
 The polytropic index N is given as, 
\begin{equation} 
\begin{aligned}
 N=\rho\frac{\partial h}{\partial p}-1=\frac{\partial f}{\partial\Theta} =6\left[(2-\xi)\frac{9\Theta^2+24\Theta/\tau+8/\tau^2}{(6\Theta+8/\tau)^2}\right] \\
 +6\xi\left[\frac{9\Theta^2+24\Theta/(\eta\tau)+8/(\eta\tau)^2}{(6\Theta+8/(\eta\tau))^2}\right]
 \label{eq:N}
 \end{aligned}
\end{equation}
And the adiabatic index is 
\begin{equation}
 \Gamma=1+\frac{1}{N}
 \label{eq:adiabatic}
\end{equation}
The sound speed is $c_s=\Gamma~\Theta$.
It can be inferred from equations \eqref{eq:N} and \eqref{eq:adiabatic} that the polytropic index and, consequently, the adiabatic index are dependent on $\Theta$ and $\xi$. Therefore, we don't have to provide them as independent variables. By examining equation \eqref{eq:adiabatic}, it becomes apparent that when the temperature is high ($\Theta \gg 1$), $\Gamma$ approaches $4/3$, and when the temperature is low ($\Theta \ll 1$), $\Gamma$ approaches 5/3.

\subsection{Steady state equations of motion} 
In the steady state $\partial/\partial t \equiv 0$. Moreover, we assume the velocity vector to have the following components $\mathbf{v}\equiv (v_r,0,v_\phi)$. 
The equations of motion are solved under these conditions.
The mass accretion rate is found by integrating the continuity equation and is given as 
\begin{equation}
\Dot{M} =-{\cal G}r^2v_r\rho
\label{eq:mass}	
\end{equation}
{where $\cal{G}= $ $4 \pi \cos{\theta_0}$, $\theta_0$ is the co-latitude (see Fig. \ref{fig:geometry})}. {Since accretion is a process where the radial velocity is directed inward so the accretion rate is negative \citep[see,][]{1999ApJ...523..340L}}. Mass accretion rate $\Dot{M}$ is a constant of motion.
Another constant of motion of the flow is obtained by integrating the momentum balance equation, and the energy equation in the steady state is given by \citep[see][]{2001ApJ...551L..77M,2004ChPhL..21.2551G,2008ApJ...677L..93B,2013MNRAS.430..386K,2014MNRAS.443.3444K},
\begin{equation}
 \epsilon = \frac{v^2_r}{2} + h - \frac{\lambda^2}{2r^2} +\frac{\lambda\lambda_0}{r^2} + \Phi(r)
\label{eq:energy}
\end{equation}
It is the specific energy {of the flow and is sometimes called as the generalized Bernoulli parameter.} It remains constant throughout the flow even in the presence of viscous dissipation, and $\lambda_0$ is the specific angular momentum at which the shear is zero presently, it is at the horizon and {$\lambda$ is the local specific angular momentum. The derivation of Eq. \ref{eq:energy} is straightforward. The second and the third terms encompass centrifugal term and the viscous transport of angular momentum of the flow. For inviscid flow, $\alpha=0$ and $\lambda=\lambda_0$, then we retrieve the expression of the Bernoulli parameter in the inviscid limit}.
Using the energy balance equation and the EoS equation given by equation \ref{eq:eos}, we can obtain the temperature gradient as 
\begin{equation}
\frac{d \Theta}{dr}= -\frac{\Theta}{N}\left(\frac{2}{r}+\frac{1}{v_r}\frac{dv_r}{dr} \right) - \frac{W^2_{r\phi}}{\rho v_r \eta_v\ N}
\label{eq:dvdtheta1}
\end{equation}
Manipulating  all the flow equations along with the EoS and using the geometric units we obtained,
\begin{equation}
\frac{dv_r}{dr}= \frac{\mathbf{Num}}{\mathbf{D}}
\label{eq:dvdr1}
\end{equation}
Where, 
\begin{equation}
\mathbf{Num} = \frac{2c^2_s}{r}+\frac{\lambda^2}{r^3} -\frac{d\Phi}{dr} +\frac{v_r\Omega_k(\lambda-\lambda_0)^2}{\alpha r^2\Theta N}
\label{eq:Num}
\end{equation}
\begin{equation}
\mathbf{D} = v_r-\frac{c^2_s}{v_r}
\label{eq:D}
\end{equation}
The azimuthal component of the momentum equation in the steady state
on integration is given by
\begin{equation}
\lambda-\lambda_0=\frac{r^2}{v_r}\frac{\alpha \Theta}{\Omega_k} \frac{d\Omega}{dr},
\label{eq:angmom}
\end{equation}
At large distances from the BH horizon, the matter is subsonic, while near the horizon, it is supersonic, therefore making it inherently transonic. As a result of this transonic behaviour, there are specific locations where $\mathbf{D}\rightarrow 0$ and $\mathbf{Num}\rightarrow 0$ in equation \eqref{eq:dvdr1}. Such a location is commonly referred to as the sonic point or critical point. Depending on flow parameters $(\epsilon,\lambda_0)$, we can have single or multiple sonic points. Again for multiple sonic points, for the particular parameters, flows can have a shock transition in between the two sonic points.
The initial point of integration is
a point asymptotically close to the horizon ($\rin =1.001~\rg$). At $\rin$, we supply a guess value of $\Theta_{\rm in}$ and compute the value of $\vin$ from given values of constants of motion $\epsilon$ and $\lambda_0$. We then integrate
equations \ref{eq:dvdr1}, \ref{eq:dvdtheta1} \& \ref{eq:angmom} simultaneously to obtained the values of $v_r$, $\Theta$ and $\lambda$. We iterate over the values of $\thein$ for a given set of $\epsilon$ and $\lambda_0$ to check for the solutions that pass through the sonic points \citep{2008ApJ...677L..93B,2014MNRAS.443.3444K}.
The solutions may be smooth or may pass through two sonic points connected by a shock.

The shock condition in an accretion disc where the viscous tensor is proportional to the shear tensor where given by several authors \cite{2008ApJ...677L..93B,2013MNRAS.430..386K,2014MNRAS.443.3444K} is essentially conservation of mass flux, momentum flux, angular momentum flux, and the energy flux,
\begin{eqnarray}
&& {\dot M}_+={\dot M}_- \\ \nonumber
&& p_++\rho_+v^2_{r+}=p_-+\rho_-v^2_{r-} \\
&& {\dot M}_+\lambda_+-r^2W_{r\phi}|_+= {\dot M}_-\lambda_--r^2W_{r\phi}|_- \\
& \& & {\dot M}_+\epsilon_+={\dot M}_-\epsilon_-
\end{eqnarray}
The suffix `$\pm$' represents the post-shock and pre-shock region of the flow, respectively.
On top of this, we assume shear less shock front ($d\Omega/dr|_+=
d\Omega/dr|_-$), which is a reasonable assumption for steady shock.
The angular momentum jump condition across the shock is given by
\begin{equation}
 \lambda_-=\lambda_++k\left[\frac{\Theta_+}{v_{r+}}-\frac{\Theta_-}{v_{r-}}\right]
 \label{eq:angjump1}
\end{equation}
Here $k=-v_{r+}(\lambda_+-\lambda_0)/\Theta_+$.
{Equation \ref{eq:angjump1} can be restructured as,
\begin{eqnarray*}
 && \lambda_--\lambda_0=(\lambda_+-\lambda_0)
 +k\left[\frac{\Theta_+}{v_{r+}}-\frac{\Theta_-}{v_{r-}}\right] \\ \nonumber
 && \lambda_--\lambda_0=(\lambda_+-\lambda_0)\frac{v_{r+}\Theta_-}{v_{r-}\Theta_+}.
\end{eqnarray*}
Since, across a shock $v_{r+}<v_{r-}$ and $\Theta_+>\Theta_-$, so $\lambda_+>\lambda_-$.}
Jump condition for 
the velocity and temperature are given by
\begin{equation}
 v^2_{r-}-2\left(k_1-h_-+\frac{\lambda_-^2}{2\rsh^2}-\frac{\lambda_-\lambda_0}{\rsh^2}\right)=0,
\end{equation}
Where
$$
k_1=\frac{v^2_{r+}}{2}+h_-\frac{\lambda_+^2}{2\rsh^2}+\frac{\lambda_+\lambda_0}{\rsh^2}
$$
\begin{equation}
 \Theta_-=k_2v_{r-}-v^2_{r-},
\end{equation}
Where 
$$
k_2=\frac{\Theta_+}{v_{r+}}+v_{r+}
$$
{The shock condition assuming hydrostatic equilibrium in the vertical direction has been computed and presented in \cite{2013MNRAS.430..386K,2014MNRAS.443.3444K}.}

\subsection{Simulation code and numerical method}
Our simulation code is implemented using the total variation diminishing(TVD) scheme, which was first introduced by \citep{1983JCoPh..49..357H}. This is a second-order accurate finite difference scheme and is Eulerian in nature. The TVD scheme has proven to be a robust approach to simulating astrophysical phenomena, as it can efficiently capture shocks. It has been widely used to study a range of astrophysical problems, including those described in \cite{2011ApJ...728..142L,2012MNRAS.423.2153C,2021MNRAS.501.4850R,2022MNRAS.509...85J,2022ApJ...933...75J}. Our code compute the spatial and temporal evolution of conserved quantities $\rho$, $\rho v_r$, $\rho v_{\theta}$, $\rho \lambda$ and $E$ using a Roe-type Riemann solver. Here $v_r$ and $v_{\theta}$ represent the velocity of the flow in r and $\theta$ direction, respectively. In most Eulerian codes without viscosity, the azimuthal momentum density ($\rho v_{\phi}$) is treated as a conserved quantity instead of the angular momentum density ($\rho \lambda$). However, in our code, we use $\rho \lambda$ instead of $\rho v_{\phi}$ to accurately compute the angular momentum such that it remains constant when there is no viscous dissipation. In a spherical coordinate system, the conserved variable($\mathbf{q}$) and the primitive variable($\mathbf{w}$) are shown in equation \ref{eq:state_primitv}. Multiplying  Eq.(\ref{eq:conserve}) by $r^2$, we obtain,
\begin{eqnarray}
&& \frac{\partial (r^2\mathbf{q})}{\partial t} + \frac{\partial (r^2\mathbf{F^r})}{\partial r}  = (r^2\mathbf{S})  \\ \nonumber
&& \frac{\partial \mathbf{\tilde q}}{\partial t}+\frac{\partial \mathbf{\tilde F}}{\partial r}
= \mathbf{\tilde S}
 \label{eq:conserve1}
 \end{eqnarray}
Here $\mathbf{\Tilde{q}}=r^2\mathbf{q}$, and accordingly the fluxes are redefined as $\mathbf{\Tilde{F}}=r^2\mathbf{F^r}$ and source term as $\mathbf{\Tilde{S}}=r^2\mathbf{S}$.
The eigenvalues of the above equation are given by,
\begin{equation}
\begin{pmatrix}
  a_1 \\
  a_2 \\
  a_3 \\
  a_4 \\
  a_5 \\
\end{pmatrix}
=
\begin{pmatrix}
 v_r-c_s \\
 v_r \\
 v_r \\
 v_r  \\
 v_r+c_s \\
\end{pmatrix}
\end{equation}
The right eigenvectors ($\mathbf{R}$) obtained are
\begin{eqnarray}
 && R_{1,5}=
 \begin{pmatrix}
  1 \\
  a_{1,5} \\
  v_\theta \\
  \lambda \\
H\mp v_r ~ c_s \\
 \end{pmatrix}
 ;~ R_{2,3}=
 \begin{pmatrix}
 0 \\
 0 \\
 1,~0 \\
 0,~1 \\
 v_\theta,~ \lambda/r^2 \\
 \end{pmatrix}  ; \\ \nonumber
 && R_4=
 \begin{pmatrix}
  1 \\
  v_r \\
  v_\theta \\
  \lambda \\
  \frac{v^2}{2}+f-N~\Theta \\
 \end{pmatrix}
\end{eqnarray}
Here,
$v^2=v_r^2+v_\theta^2+v_\phi^2$ and $H=h+0.5~v^2$. Left eigenvectors ($\mathbf{L}$) corresponding to right eigenvectors are obtained from the orthogonality condition,
\begin{equation}
 R_{i}L_{j}=\delta_{ij}
 \label{eq:orth}
\end{equation}
Where $R_{i}$ and $L_{j}$ are the component of the right and left eigenvectors respectively.
The updating of state vector $\mathbf{\Tilde{q}}^n$ to $\mathbf{\Tilde{q}}^{n+1}$ follows the procedure of \citep{1983JCoPh..49..357H},
\begin{equation}
\mathbf {\tilde{q}}_i^{n+1}=\mathbf{\Tilde{q}}_i^n-\frac{\Delta t^n}{\Delta x}\left(\Bar{f}_{x,i+1/2}-\Bar{f}_{x,i-1/2}\right),
 \label{eq:lx}
\end{equation}

\begin{equation}
\Bar{f}_{x,i+1/2}= \frac{1}{2}[\mathbf{\Tilde{F}}(\mathbf{\Tilde{q}}^n_i) +\mathbf{\Tilde{F}}(\mathbf{\Tilde{q}}^n_{i+1})] -\frac{\Delta x}{2 \Delta t^n}  \sum_{k=1}^{5} \beta_{k,i+1/2} \mathbf{R}^n_{k,i+1/2},
 \label{eq:fi}
\end{equation}

\begin{multline}   
\beta_{k,i+1/2}=Q_k \left( \frac{\Delta t^n}{\Delta x} a^n_{k,i+1/2}+\gamma_{k,i+1/2} \right)\alpha_{k,i+1/2} \\
-(g_{k,i}+g_{k,i+1}),
 \label{eq:beta}
 \end{multline}

\begin{equation}
\alpha_{k,i+1/2}= \mathbf{L}^n_{k,i+1/2}(\mathbf{\Tilde{q}}^n_{i+1}-\mathbf{\Tilde{q}}^n_{i-1}),
 \label{eq:alfa}
\end{equation}

\begin{equation}
 \gamma_{k,i+1/2}=\begin{cases}
    \left( g_{k,i+1}-g_{k,i}\right)/\alpha_{k,i+1/2}, & \text{if $\alpha_{k,i+1/2}\not=0$}\\
    0, & \text{if $\alpha_{k,i+1/2}=0$}
  \end{cases}
   \label{eq:gammas}
\end{equation}

\begin{multline}
  g_{k,i}=sign(\tilde{g}_{k,i+1/2}) \max\{0,\min[\lvert \tilde{g}_{k,i+1/2} \lvert, \\
   sign(\tilde{g}_{k,i+1/2})\tilde{g}_{k,i+1/2}]\},
 \label{eq:g_ik}
 \end{multline}

\begin{equation}
 \tilde{g}_{k,i+1/2}=\frac{1}{2}\left[ Q_k\left(\frac{\Delta t^n}{\Delta x} a^n_{k,i+1/2}\right)-\left(\frac{\Delta t^n}{\Delta x} a^n_{k,i+1/2}\right)^2\right]\alpha_{k,i+1/2},
   \label{eq:g_tilde}
\end{equation}

\begin{equation}
 Q_k(x)=\begin{cases}
    x^2/4\xi_k+\xi_k, & \text{if $\lvert x \lvert< 2\xi_k$}\\
    \lvert x \lvert, & \text{if $\lvert x \lvert> 2\xi_k$}
  \end{cases}
   \label{eq:Q_x}
\end{equation}
Here $k=1-5$ goes for the five characteristic modes. The parameters $\xi_k$ implicitly control numerical viscosity, and they are defined for $0\leq \xi_k \leq 0.5$.
The updated fluid state variables $\mathbf{q}$ are obtained by incorporating the source terms with $\mathbf{\tilde q}$s
\begin{equation}
\mathbf{q}_i^{n+1}=\frac{1}{r^2}\left(\mathbf{\Tilde{q}}_i^{n+1}+ \Delta t^n \mathbf{\Tilde{S}}(\mathbf{\Tilde{q}}_i^n)\right)
 \label{eq:source}
\end{equation}
The upgraded primitive variables ($\mathbf{w}$) are computed from the $\mathbf{q}$.
Recalling the Eq. \ref{eq:state_primitv}, 
The relations between the first four components of state and primitive variables are easy
\begin{eqnarray}
&& w_1=\rho=q_1;~ w_2=v_r=q_2/q_1=M_r/\rho; \\ \nonumber
&& w_3=v_\theta=q_3/q_1=M_\theta/\rho; \\ \nonumber
&& w_4=\lambda=q_4/q_1=M_\lambda/\rho
\label{eq:primit2}
\end{eqnarray}
The relation between $w_5$ and $q_5$ is not straightforward.
From Eq. \ref{eq:state_primitv} we know $E=\rho v^2/2 +e$ and combining the expression of EoS (Eq. \ref{eq:eos}), we obtain,
\begin{equation}
 f=\frac{E}{\rho} -\frac{v^2}{2}=C
 \label{eq:E}
\end{equation}
Here $C$ is known at each of the computations cells from the updated values of $E$, $\rho$, and $v^2$.
Combining Eq.(\ref{eq:f}) with  Eq.(\ref{eq:E}), we obtain the cubic equation of $\Theta$ as,
\begin{equation}
 27\Theta^3+9K_1\Theta^2+12K_2\Theta+\frac{16(1-C)}{\eta \tau^2}=0,
 \label{eq:cubic}
\end{equation}
where,
\begin{multline*}
 K_1=\frac{(1+2/\eta)(2-\xi)+\xi(1/\eta+2)}{\tau}+(1-C), \\
 K_2=\frac{2}{\eta \tau^2} +\frac{(1-C)}{\tau}(1+1/\eta)
\end{multline*}
Eq.(\ref{eq:cubic}) has an analytic solution. We follow the standard methods described by  \cite{1970hmfw.book.....A} to solve the cubic equation and found that the CR EoS admits only one unique root. Defining $b_1=9K_1/27$; $b_2=12K_2/27$; and $b_3=16(1-C)/(27\eta \tau^2)$, then

\begin{gather*}
Q=(3b_2-b^2_1)/9; \\
T=(9b_1b_2-27b_3-2b^3_1)/54
\end{gather*}
and a discriminant, $\Xi=Q^3+T^2$.
\\
The number of roots of Eq. (\ref{eq:cubic}) will depend on the value of $\Xi$. In the case of CR EoS, we have $\Xi<0$, and $Q<0$, which means the cubic equations has three real but unequal roots. The physical root of Eq.(\ref{eq:cubic}) is,
\begin{equation}
 \Theta=2\sqrt{-Q}\cos(\frac{\Tilde{\theta}}{3})-\frac{b_1}{3}
 \label{eq:root}
\end{equation}
Where, $\cos(\Tilde{\theta})=\frac{T}{\sqrt{-Q^3}}$.
\\ 
Once we get $\Theta$, we retrieve the pressure from the definition of $\Theta$ i.e. $p=\rho \Theta$.

The computational box is divided into equal radial grids. The incoming matter enters the computational box through the outer boundary. 
We set the density of the incoming gas, denoted as $\roi$, to a value of 1. This choice is made under the assumption of no cooling or self-gravity, allowing for the density to be scaled out of the calculations. At the outer boundary, we provide the radial velocity $v_r$ and the temperature $\Theta$ of the flow. These values are obtained from analytical results. To simulate the behaviour near the black hole horizon, we have introduced an absorbing inner boundary situated at $1.5\rg$. Within this boundary, all material is fully absorbed into the black hole. The background material in the simulation is characterized by a density of $\robg = 10^{-6}$. Pressure at the outer boundary and the background material has been calculated from $\Theta$.

\section{Code Verification}
 Two tests have been presented to demonstrate that the code can handle the transonic flows in the vicinity of black holes. We test the code with the steady-state analytic accretion solution 
 around a black hole in spherical geometry, both for the inviscid flow and flows in the presence of viscosity.
\begin{figure}
	\centering
	\includegraphics[width=2.3 in]
	{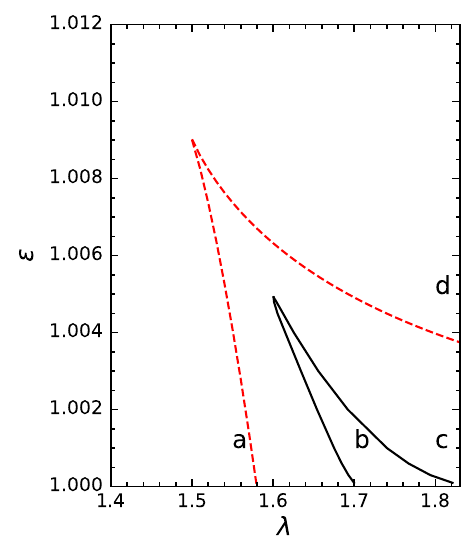}
        \caption{The domain of multiple sonic points (dashed) and shock (solid) in $\epsilon-\lambda$ Parameter space in the inviscid accretion flows.}
        \label{fig:parameter}
\end{figure}
        
\begin{figure}
        \centering
        \includegraphics[width=3 in]
        {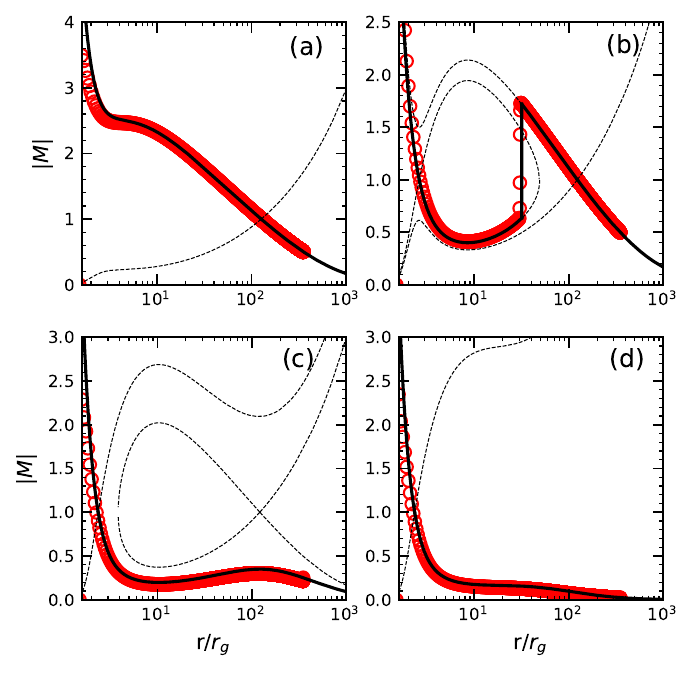}
        \caption{Comparison of numerical simulation (red open circle) and the analytic solution (black solid line) for a one-dimensional inviscid accretion flow in spherical geometry. The absolute value of Mach number (M) as a function of $r$. The injection radius or the outer boundary is $\rou=350$ for all the cases. For panels (a)$~\vou=-1.722\times 10^{-2},~\theou=7.869\times 10^{-4}$ (b) $~\vou=-1.706\times 10^{-2},~\theou=7.871\times 10^{-4}$ (c) $\vou= -8.768\times 10^{-3},~\theou=8.215\times 10^{-4}$, and (d)
        $\vou=-1.336\times 10^{-3},~\theou=2.099\times 10^{-3}$. The simulation is conducted over 4096 uniform grid cells.}
        \label{fig:inviscid}      
\end{figure}

\subsection{Inviscid flow} 
In this section, we compare the theoretically obtained steady-state solutions with the results of the simulation code in the absence of viscosity. 
In the absence of viscosity, angular momentum is constant throughout the flow. For strong gravity, if the flow angular momentum $\lambda < \lambda_{\rm ms}$  then accretion is possible. Here $\lambda_{\rm ms}$ is the specific angular momentum of a particle rotating in ISCO or the marginally stable orbit. For the inviscid case, we choose $\lambda<\lambda_{\rm ms}$. The classification of the $\epsilon-\lambda$ parameter space was first presented by \cite{1987PASJ...39..309F,1989ApJ...347..365C}.
If the flow angular momentum is low, then the solution may pass through a single sonic point located far from the central object, and the solution will be Bondi type. However, for low $\lambda$ but higher values of $\epsilon$, the sonic points are located closer to the horizon. For flows with higher or intermediate $\lambda$, the flow may harbour multiple sonic points. With changing angular momentum,
the spherical symmetry imposed by gravity is modified due to the
presence of rotation, which causes the formation of multiple sonic points. If the rotation is strong enough,
then it can trigger shock transitions in accretion flows.
In Fig. \ref{fig:parameter} the domain which admits multiple sonic points {or multiple critical points} (abbreviated as MCP) in the $\epsilon-\lambda$ parameter space is marked with the red dashed curve. While the shock parameter space is marked by the black solid curve. On this parameter space, we marked four regions in the $\epsilon$-$\lambda$ parameter space as `a' ($\epsilon,~ \lambda ~ \equiv ~ 1.001,~1.55$), `b' ($\epsilon,~ \lambda ~ \equiv ~1.001 ,~1.72$), `c' ($\epsilon,~ \lambda ~ \equiv ~ 1.001,~1.8 $), and `d' ($\epsilon,~ \lambda ~ \equiv ~1.005 ,~1.8$).
In Fig. \ref{fig:inviscid}a-d, we compare the theoretical solutions (absolute values of Mach number $M=v_r/c_s$ with $r$) for the flow parameters corresponding to the points a, b, c, and d points of Fig. \ref{fig:parameter}. The simulation
result is obtained by injecting the values of flow variables at an injection radius $\rou$ taken from the analytical solution at that point.
{Since for accretion $v_r<0$, so $M$ is also negative, and that is why we plot absolute value of Mach number or $|M|$.} In Fig. \ref{fig:inviscid}a, we compare the theoretical accretion curve (black solid) with the simulation result (red open-circle) corresponding to point `a' of Fig. \ref{fig:parameter}. The dotted line is an excretion-type flow that fulfills the outflow-type boundary condition. 
The injection radius for this simulation is also the outer boundary of simulation $\rou=350$, and injection flow variables at the $\rou$ are $\vou=-1.722\times 10^{-2},~\theou=7.869\times 10^{-4}$.
The numerical simulation result follows the analytical curve closely and in fact, it becomes transonic around the
analytical value of $\rco$. The theoretically obtained sonic point is at $\rco=126.414$ and is located at the crossing point of the dotted and solid black curves, while the sonic point obtained through numerical simulations is at $126.425$, a discrepancy of only $0.0087\%$.
In Fig. \ref{fig:inviscid}b, we compare the analytical accretion (black, solid) with the simulation result (red open-circle) corresponding to the point `b' of Fig. \ref{fig:parameter}. The injection parameters are $\rou=350,~\vou=-1.706\times 10^{-2},~\theou=7.871\times 10^{-4}$. This solution corresponds to higher $\lambda$ and admits multiple sonic points. The flow becomes supersonic through the outer sonic point. The supersonic flow then `feels' the rotational barrier due to higher values of $\lambda$.
This barrier along with the thermal pressure, acts as a barrier to the supersonic flow and discontinuously slows down the flow in the form of a shock or $\rsh= 31.83$ (vertical jump) and jumps to the subsonic branch. The flow again becomes supersonic and dives into the BH after passing through the inner sonic point ($\rci=2.617$). Dotted curves present all possible solutions through the two physical sonic points available, but the physically correct accretion solution is the one traced by the simulation (red open circles). In simulations, the shock is captured in 6 cells. The analytically determined outer sonic point is at
$\rco=123.306$, while in the simulations it is at $123.323$, a fractional difference of only $0.0138\%$. The analytical inner sonic point is at $\rci=2.617$ while numerically it is at $2.667$, a mere $1.9 \%$ in fractional change. The error is more in locating the inner sonic point however, it is indeed quite low.
In Figs. \ref{fig:inviscid}c \& d we compare analytical and simulation results for flow parameters corresponding to points `c' and `d' of Fig. \ref{fig:parameter}. Since the flow in these two panels is of even higher angular momentum, the accretion flow velocity remains largely subsonic and eventually becomes supersonic to dive into the BH through the inner type sonic point $\rci=2.511$ (for solution corresponding to point c ) and $\rci=2.357$ (solution corresponding to the point d). The injection parameters are $\vou=- 8.768\times 10^{-3},~\theou=8.215\times 10^{-4}$, and $\vou=-1.336\times 10^{-3},~\theou=2.099\times 10^{-3}$, respectively. In both these cases, inner sonic points obtained with our simulation code matched the analytical result at around $2\%$ accuracy.
It may be noted that in the case of Fig.
\ref{fig:inviscid}(c), we did not inject with the flow variables of $\alpha$ type topology. The main reason is that the $\alpha$ type topology is not global, but even if we choose to inject with flow values corresponding to the $\alpha$ type branch of Fig.
\ref{fig:inviscid}(c), the injected matter jumps to the global solution
(the black, solid curve) and closely follows it.

\begin{figure}
	\centering
	\includegraphics[width=2.3 in]{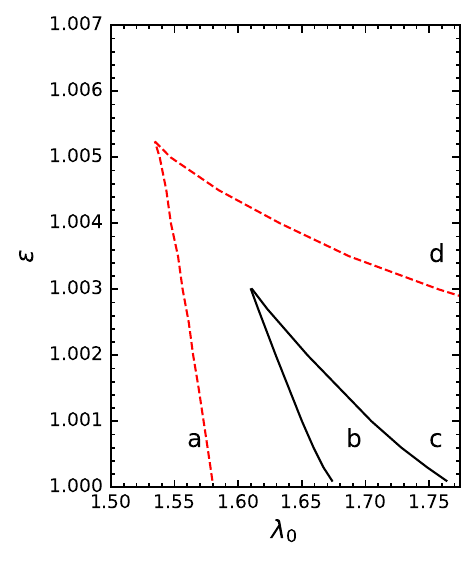}
        \caption{The domain of multiple sonic points (dashed) and shock (solid) in $\epsilon-\lambda$ Parameter space for viscous flows ($\alpha=0.01$).}
        \label{fig:parameter_visc}
\end{figure}

\begin{figure}
        \centering
        \includegraphics[width=3 in]{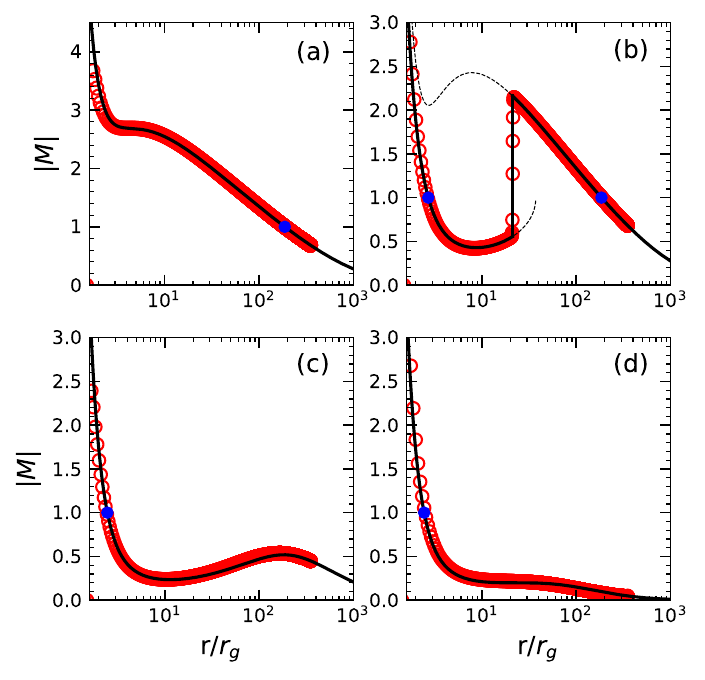}
        \caption{Comparison of $M$ vs $r$ from numerical simulation (red, open circle) with the analytic solution (black, solid). The viscosity parameter is $\alpha=0.01$. The injection values are $\rou=350$ for all the cases. For panels (a)$~\vou=-2.102\times 10^{-2},~\theou=6.322\times 10^{-4},~\lou=1.590$ (b) $~\vou=-2.091\times 10^{-2},~\theou=6.324\times 10^{-4},~\lou=1.7212$ (c) $\vou= -1.419\times 10^{-2},~\theou=6.709\times 10^{-4},~\lou=1.794$, and (d) $\vou=-2.317\times 10^{-3},~\theou=1.598\times 10^{-3},~\lou=2.431$. 4096 uniform grid cells are used in the simulation. The blue dot represents the sonic points.}
        \label{fig:viscous}      
\end{figure}

\begin{figure}
        \centering
        \includegraphics[width=3 in]{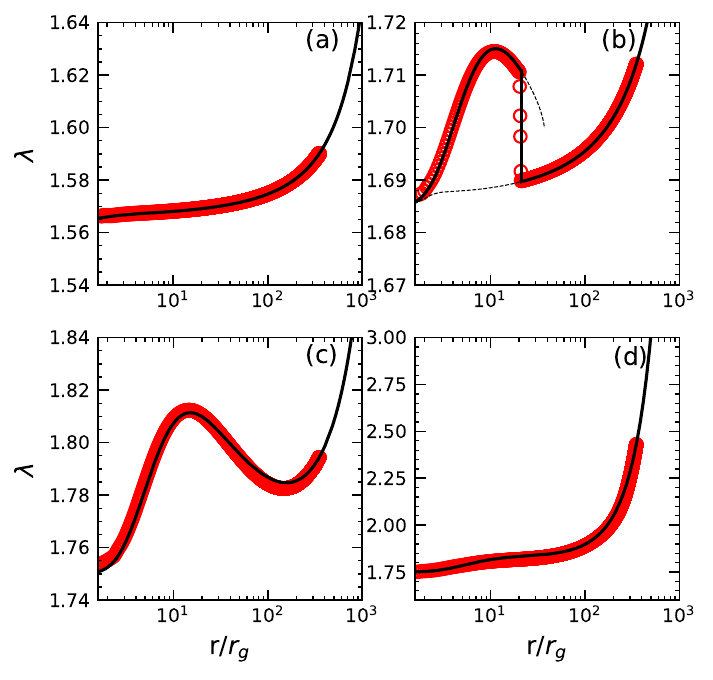}
        \caption{Comparison of $\lambda$ vs $r$ from numerical simulation (red open circle) and the analytic solution (black solid). The viscosity coefficient is $\alpha=0.01$. This figure corresponds to Fig. \ref{fig:viscous}. The numerical solution is obtained using 4096 uniform grid cells.}
        \label{fig:viscous_l}      
\end{figure}

\subsection{Viscous flow} 
\subsubsection{Same viscosity parameter different $\epsilon$ and $\lambda_0$}
In this section, we compare the theoretical steady-state, viscous solutions with the results from the simulation code and see how our code manages to tackle viscous dissipation. 
Viscosity redistributes and transports the angular momentum outwards and, in the process, heats up the flow. So, a code should be able to capture the angular momentum distribution accurately in order to understand the accretion disc dynamics properly.
It may be noted that in the presence of viscosity, the generalized Bernoulli parameter $\epsilon$ (Eq. \ref{eq:energy}) is a constant of motion, while $\lambda_0$ is the constant of integration, therefore the parameter space for viscous accretion is marked in the $\epsilon-\lambda_0$.
In Fig. \ref{fig:parameter_visc}, we mark the domain that admits multiple sonic points or the MCP region with a red dashed curve for the viscosity parameter
$\alpha=0.01$. The shock parameter space is marked by the black solid curve. Due to the reduction of the angular momentum along the flow in the presence of viscosity, it causes a shift in the MCP and shock parameter space toward the lower end of the $\lambda_0$ scale, resulting in a narrower range. The cusp of the MCP region for inviscid flow is around $\epsilon \sim 1.009$
(see, Fig. \ref{fig:parameter}), while in Fig. \ref{fig:parameter_visc}, the cusp is at $\epsilon=1.0052$. Similarly, the shock region is also smaller. 
We marked four points in the $\epsilon$-$\lambda_0$ parameter space as a ($\epsilon,~ \lambda_0 ~ \equiv ~ 1.0006,~1.565$), b ($\epsilon,~ \lambda_0 ~ \equiv ~1.0006 ,~1.685$), c ($\epsilon,~ \lambda_0 ~ \equiv ~ 1.0006,~1.75 $), and d ($\epsilon,~ \lambda_0 ~ \equiv ~1.0034 ,~1.75$). Points a, c, and d represent regions that admit global solutions passing through only one sonic point, respectively. While point b represents the region that admits steady shock. In Fig. \ref{fig:viscous}a-d, we compare the theoretical semi-analytical solutions ($|M|$) with flow parameters corresponding to the points a, b, c, and d of Fig. \ref{fig:parameter_visc}. 
The injection radius used in the simulation code for all the plots in Figs. \ref{fig:viscous} \&
\ref{fig:viscous_l} is $\rou=350$. In Fig. \ref{fig:viscous}a, we compare the theoretical curve (black solid) with the simulation result (red open-circle) corresponding to the point `a' of Fig. \ref{fig:parameter_visc}. The injection flow parameters are $~\vou=-2.102\times 10^{-2},~\theou=6.322\times 10^{-4},~\lou=1.59$. The flow is for low angular momentum 
and due to viscosity, angular momentum transport will be such that the solution passes through an outer type single sonic point
$\rco=187.15$ (indicated by the thick blue dot). 

In Fig. \ref{fig:viscous}b, we compare the analytical (black solid) with the simulation result (red open-circle) corresponding to the point `b' of Fig. \ref{fig:parameter_visc}. The injection parameters are $~\vou=-2.091\times 10^{-2},~\theou=6.324\times 10^{-4},~\lou=1.7212$. This solution corresponds to higher $\lambda_0$ and admits multiple sonic points. The flow becomes supersonic through the outer sonic point ($\rco=184.89$). The supersonic flow then `feels' the rotational barrier due to higher values of $\lambda$. This, in addition to the thermal pressure, acts as a barrier to the supersonic flow and discontinuously slows down the flow in the form of a {stationary} shock {at the radial distance} $\rsh=21.18$ (vertical jump) and jumps to the subsonic branch. The flow again becomes supersonic and dives into the BH after passing through the inner sonic point ($\rci=2.71$). The dotted curve through $\rco$ presents the part of the supersonic branch through which the matter would have accreted if there was no shock. Again, the shock is resolved in 6 cells. {Accretion shocks, if driven by the $r - \phi$ component of the viscous stress tensor, are thin shocks. However, if the viscosity originates from the $\theta - \phi$ component or $r-\theta$ component of the stress tensor, then the shock front might be of a finite width \citep{1966egct.book.....Z}. Since $r - \phi$ component of the viscous stress tensor is dominant, the type of shock we study in this work is thin}.
In Figs. \ref{fig:viscous} c \& d we compare analytical and simulation results for flow parameters corresponding to points `c' and `d' of Fig. \ref{fig:parameter_visc}. Since the flow in these two panels is for even higher angular momentum, the accretion flow velocity remains largely subsonic and eventually becomes supersonic to dive into the BH through the inner type sonic point $\rci$. The injection parameters are, $\vou= -1.419\times 10^{-2},~\theou=6.709\times 10^{-4},~\lou=1.794$, and $\vou=-2.317\times 10^{-3},~\theou=1.598\times 10^{-3},~\lou=2.431$, respectively.

As we are examining viscous accretion flows, it is necessary to ensure that our code can handle the transport of angular momentum. In Fig. \ref{fig:viscous_l}a-d, we have compared the theoretical $\lambda$ distribution (solid, black) with the simulation (red, open circle) corresponding to the solutions
presented in Fig. \ref{fig:viscous}a-d. 
{At the shock, the $\lambda$ jumps to a higher value in the post-shock flow, as is clear from
Eq. \ref{eq:angjump1}}.  Here also, wherever $|M|$ dips, which implies hotter and slower flow,  
Most interestingly, in Fig. \ref{fig:viscous_l}b, the simulations show that both the post-shock branch through $\rci$ and the pre-shock branch through $\rco$ in the shocked solutions are solutions with the same $\lambda_0$. In \cite{ 2013MNRAS.430..386K}, we proposed that $\lambda_0$ for the branches would be the same and obtained the solution, in this paper, we show that
even numerical simulations confirm our proposal.
{The solution represented by Figure \ref{fig:viscous}c
is not a shock solution, it is a smooth but not a monotonically increasing or decreasing function. The angular momentum piles up in regions where the magnitude of the Mach number is low. It implies that, if the flow is slow ($|M|<1$) and hot ($\Theta$ high),
viscous transport of $\lambda$ is significant
(Fig. \ref{fig:viscous_l}c).}

\begin{figure*}
	\begin{center}
       \includegraphics[width=6 in]{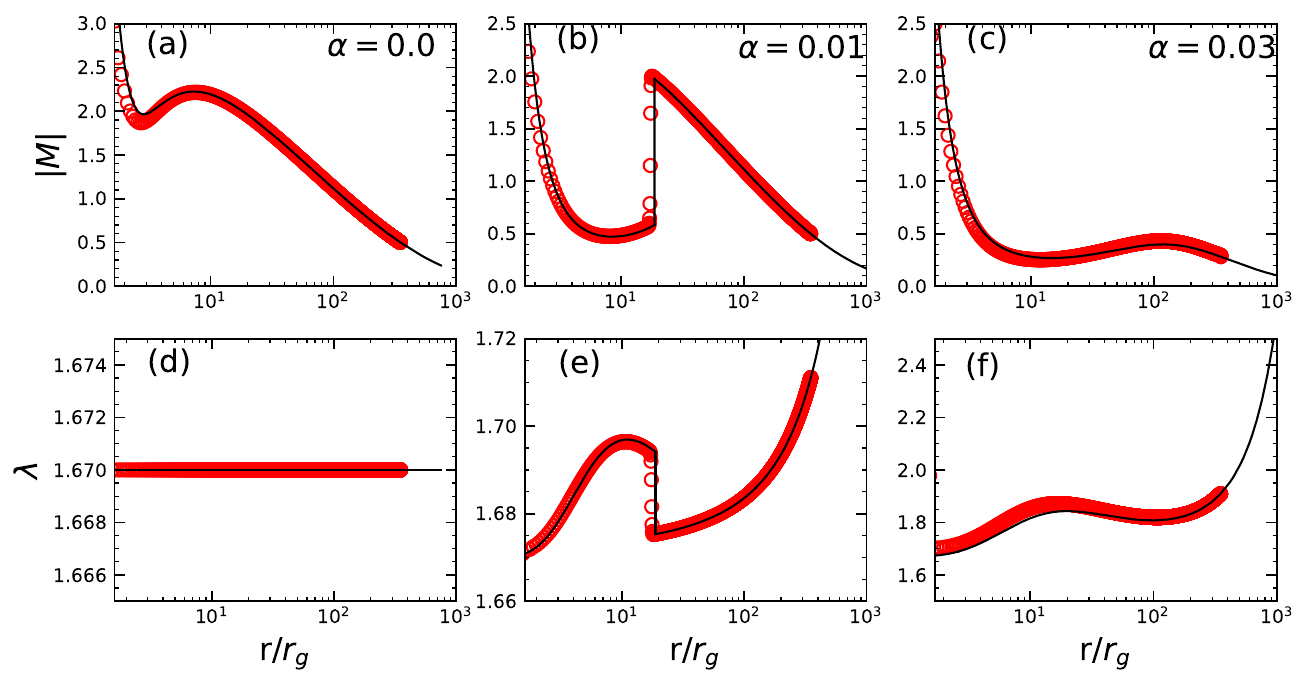}
        \caption{Mach number (panels a, b, c) and angular momentum (d, e, f) variation with $r$
        (in units of $\rg$) for $\epsilon=1.0005$ and $\lambda_0=1.67$. Viscosity increased from left to right,
        $\alpha=0$ (a, d), $\alpha=0.01$ (b, e) and $\alpha=0.03$
        (c, f). $\epsilon$ and $\lambda_0$ are taken to be 1.0005, 1.67. The injection values are $\rou=350$ for all the cases. For panels (a) $~\vou=-0.0171,~\theou=0.7870\times 10^{-3}, ~\lou=1.67$ (b) $~\vou=-0.0171, ~\theou=0.7872\times 10^{-3}, ~\lou=1.7111$, and (c) $~\vou=-0.0097, ~\theou=0.8190\times 10^{-3}, ~\lou=1.9092$. }
        \label{fig:same_inner}
        \end{center}
\end{figure*}

\subsubsection{Same $\epsilon$ and $\lambda_0$ but different $\alpha$}
We would like to study how viscosity affects the flow variables for given constants of motion.
\cite{2013MNRAS.430..386K} analytically showed that by keeping $\epsilon$ and $\lambda_0$ same if we increase viscosity parameter ($\alpha$) solution topology changes. This amounts to studying accretion flows with the same inner boundary condition. \citeauthor{2013MNRAS.430..386K} showed in their theoretical study that an accretion flow with $\alpha=0$, which is flowing through an outer sonic point and does not have shock, may undergo shock transition with the increase of $\alpha$. On the other hand, solutions with shocks in inviscid limit 
will become smooth solutions with the increase of viscosity.
All these depend on the angular momentum transport due to viscosity.
In this paper, we show, through simulations, one case where a solution has no shock in the inviscid regime but generates shock as viscosity is increased, and with further increase in the viscosity parameter, the solution becomes shock-free, all the time keeping $\epsilon$ and $\lambda_0$ same. {Since $\epsilon$ and $\lambda_0$ is same for all the solutions, but
$\alpha$ is different for each of them. Therefore, the outer boundary conditions of all three cases presented in Figs. \ref{fig:same_inner} are different.}  
%

We compare analytical accretion solutions for flow parameters $\epsilon=1.005$ and $\lambda_0=1.67$ with numerical simulation solutions. For the numerical simulation, we injected flow variables obtained from the analytical solution. The $\rou=350$ and the code had $3072$ uniform grids. The injection parameters are mentioned in the caption of Fig.
\ref{fig:same_inner} {and from the panels left to right, the injection speed, temperature, and angular momentum increase with increasing $\alpha$}. Black solid lines represent analytic accretion solutions, whereas red open circles show the simulated solutions. In Fig. \ref{fig:same_inner}a \& d, we plot the Mach number $M$ and the
specific angular momentum $\lambda$ as a function of $r$
for inviscid ($\alpha=0$) accretion flow. In Fig. \ref{fig:same_inner}b \& e we plot the accretion solution with $\alpha=0.01$, and the accretion solution for $\alpha=0.03$ are plotted in Fig. \ref{fig:same_inner}c \& f.
Values of injection parameters, i.e., velocity, temperature, and angular momentum, are taken from the analytic solution. 
In the lower panels (d, e, f), the $\lambda$ distribution is shown, and $\lambda$ is strictly constant for inviscid flow or $\alpha=0.0$, and for the viscous flow, the angular momentum is no more constant
but the simulation code quite accurately computes it.
There is a shock in the accretion disc at $18.8r_g$ for $\alpha=0.01$ from the analytic result. But from the simulation, we get a shock at around {\textbf{$17.37r_g$}}.
For this set of flow constants $\epsilon,~\lambda_0$, we checked that
analytically, accretion flow will harbour a steady shock if $0.008 \leq \alpha
\leq 0.0279$.
But in simulation, we find shock in the range $0.009 \leq \alpha \leq 0.0285$. {It may be noted that we only compared the steady semi-analytical accretion solution for the same inner boundary condition. There is no shock in Fig. \ref{fig:same_inner} (c \& f) for $\alpha=0.03$. Since, at the outer boundary, the flow variables have very different values therefore it is not wise to conclude at $\alpha=0.03$, there is no shock. The angular momentum at the outer boundary i. e., $\lambda_{\rm ou}$
is high, so the magnitude of infall velocity $v_r$ is low, and the flow remains subsonic till very close to the black hole. 
Therefore, there is no shock.}

\section{Results: Oscillating shocks}
\label{sec:results}
\begin{table}
\centering
    \caption{Details of injection parameters for different models.}
    \label{tab:injection}
    \begin{tabular}{llccl}
    \hline
    Model & $\rou$  & $\vou$ & $\theou$ & $\lou$ \\
       & ($r_g$) & (c) & &($r_g$ c) \\
    \hline
    \hline
    L1 & 1000   &  $-0.1348\times 10^{-1} $    & $0.1912\times 10^{-3} $ & 1.76 \\
    L2 & 1000   &  $-0.1348\times 10^{-1}$    & $0.1912\times 10^{-3}$  &  1.77 \\
    L3 &  1000  & $-0.1348\times 10^{-1}$    &  $0.1912\times 10^{-3}$ & 1.78\\
     E1 &  1000  & $-0.1539\times 10^{-1}$  & $0.1632\times 10^{-3}$ & 1.78\\
    E2 & 1000   &  $-0.1348\times 10^{-1}$   & $0.1912\times 10^{-3}$ & 1.78 \\
    E3 & 1000 &  $-0.1215\times 10^{-1}$    & $0.2152\times 10^{-3}$ & 1.78 \\
    \hline
    \end{tabular}   
\end{table}
\begin{figure*}
       \includegraphics[width=6 in]{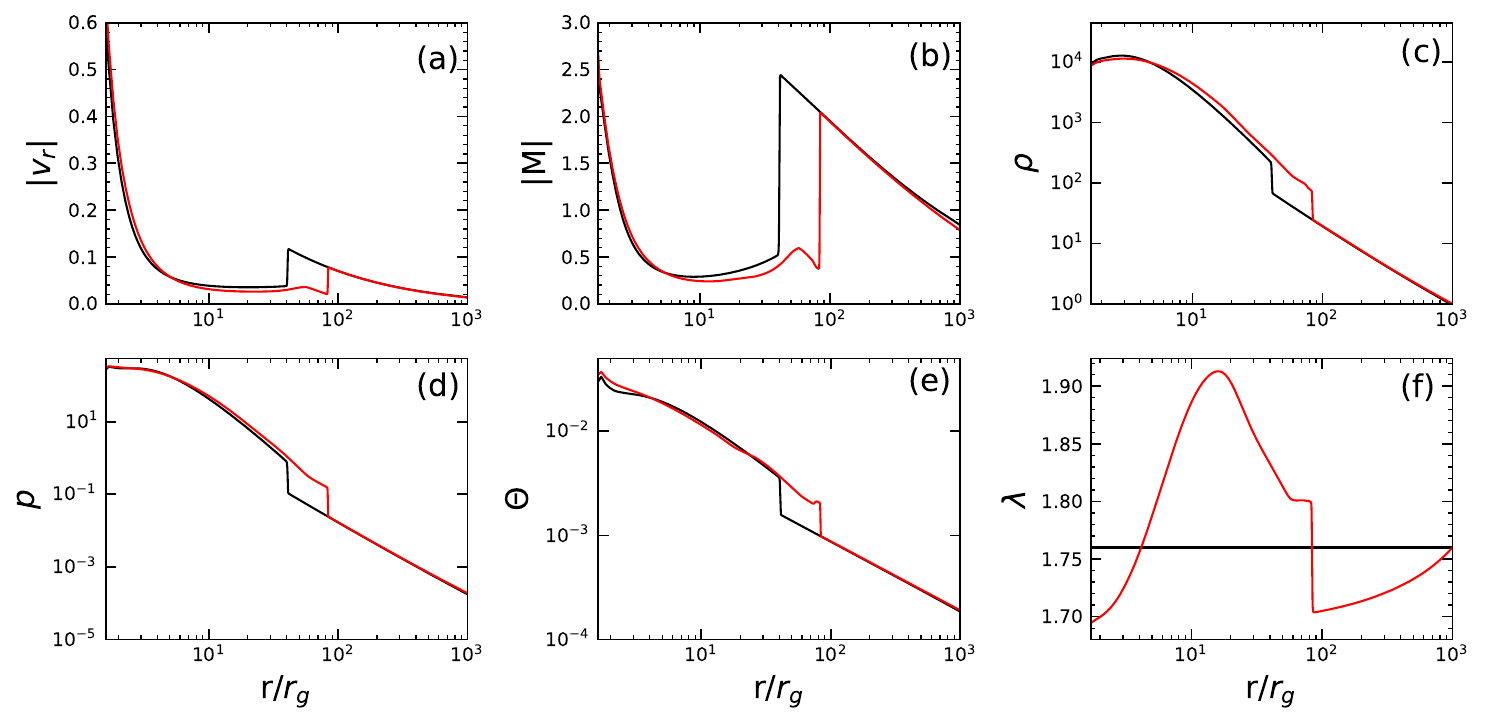}
        \caption{{{Snapshot of flow variables (a) $v$, (b) $M$
        (c) log$(\rho)$, (d) log$(p)$, (e) log$(\Theta)$, and
        (f) $\lambda$ at times $t_1=200$ and $t_2=380$ in code unit. Flow variables at $t_1=200$ present the steady state of inviscid accretion (shown in black color), and $t_2=380$ present the dynamic state of viscous accretion where the shock is oscillating (shown in red color) for model L1. The viscosity parameter is $\alpha$=0.035.}}}
        \label{fig:variables}
\end{figure*}
\begin{figure}
	\begin{center}
       \includegraphics[width=3.3 in]{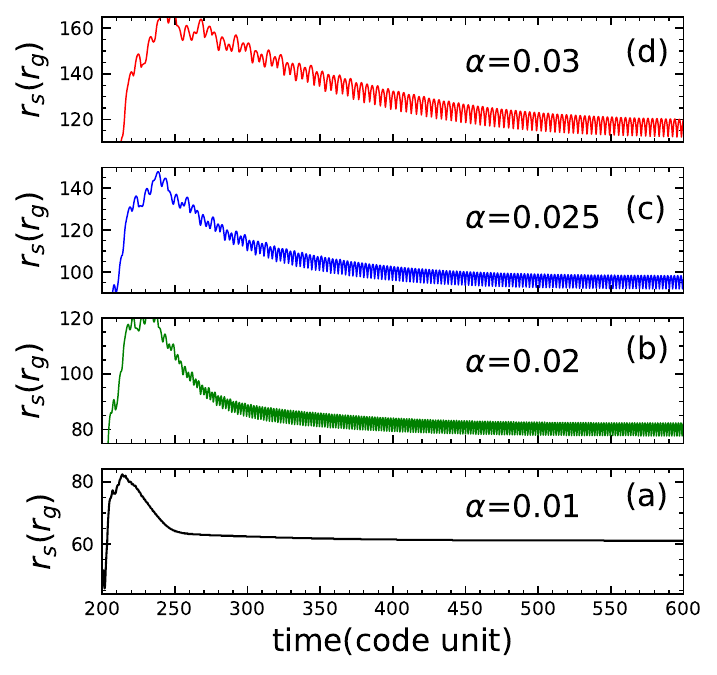}
        \caption{{{Time evolution of shock position for model L2. Viscosity parameters are (a) $\alpha=0.01$, (b) $\alpha=0.02$, (c) $\alpha=0.025$ and (d) $\alpha=0.03$.}}}
        \label{fig:different_alpha}
        \end{center}
\end{figure}
\begin{figure*}
       \includegraphics[width=6 in]{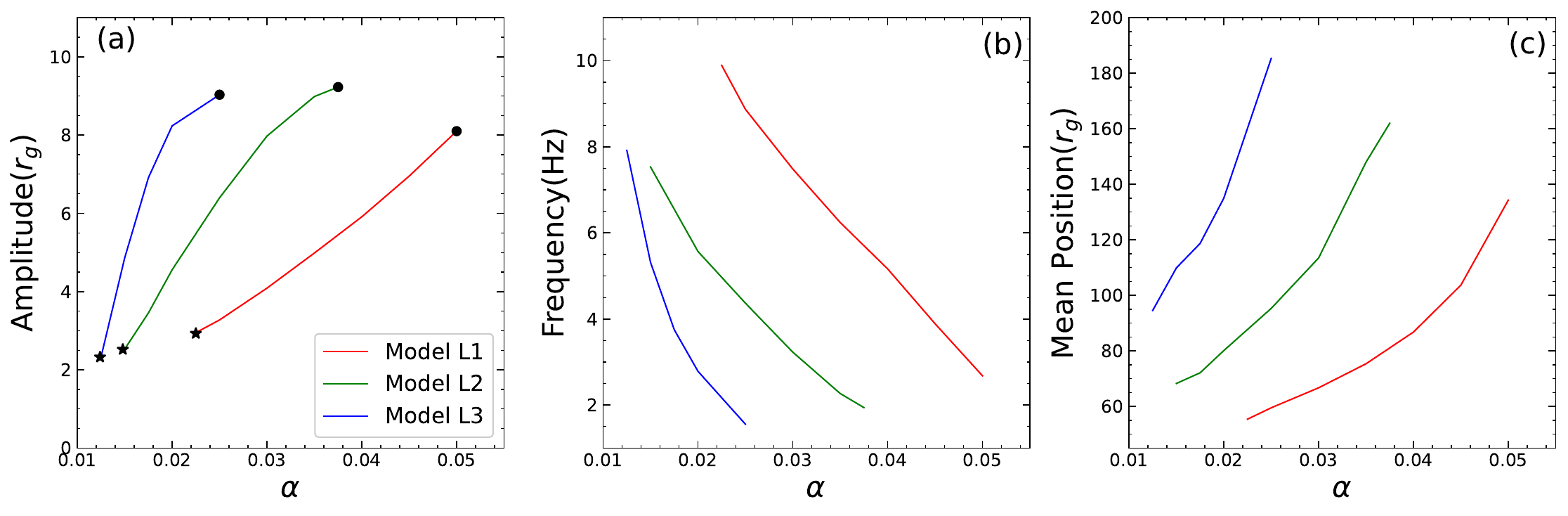}
    \caption{{{Variation of amplitude (a), frequency of the oscillation (b), and the mean position of the shock (c) with the viscosity parameter ($\alpha$) assuming black hole mass to be $\mbh=10M_{\odot}$. Flow parameter $\epsilon$=1.0001. Red line is for $\lou=1.76$ (model L1), green line is for $\lou=1.77$ (model L2), blue line is for $\lou=1.78$ (model L3).}}}
        \label{fig:alfa_var}
\end{figure*}

\begin{figure*}
	\begin{center}
       \includegraphics[width=6 in]{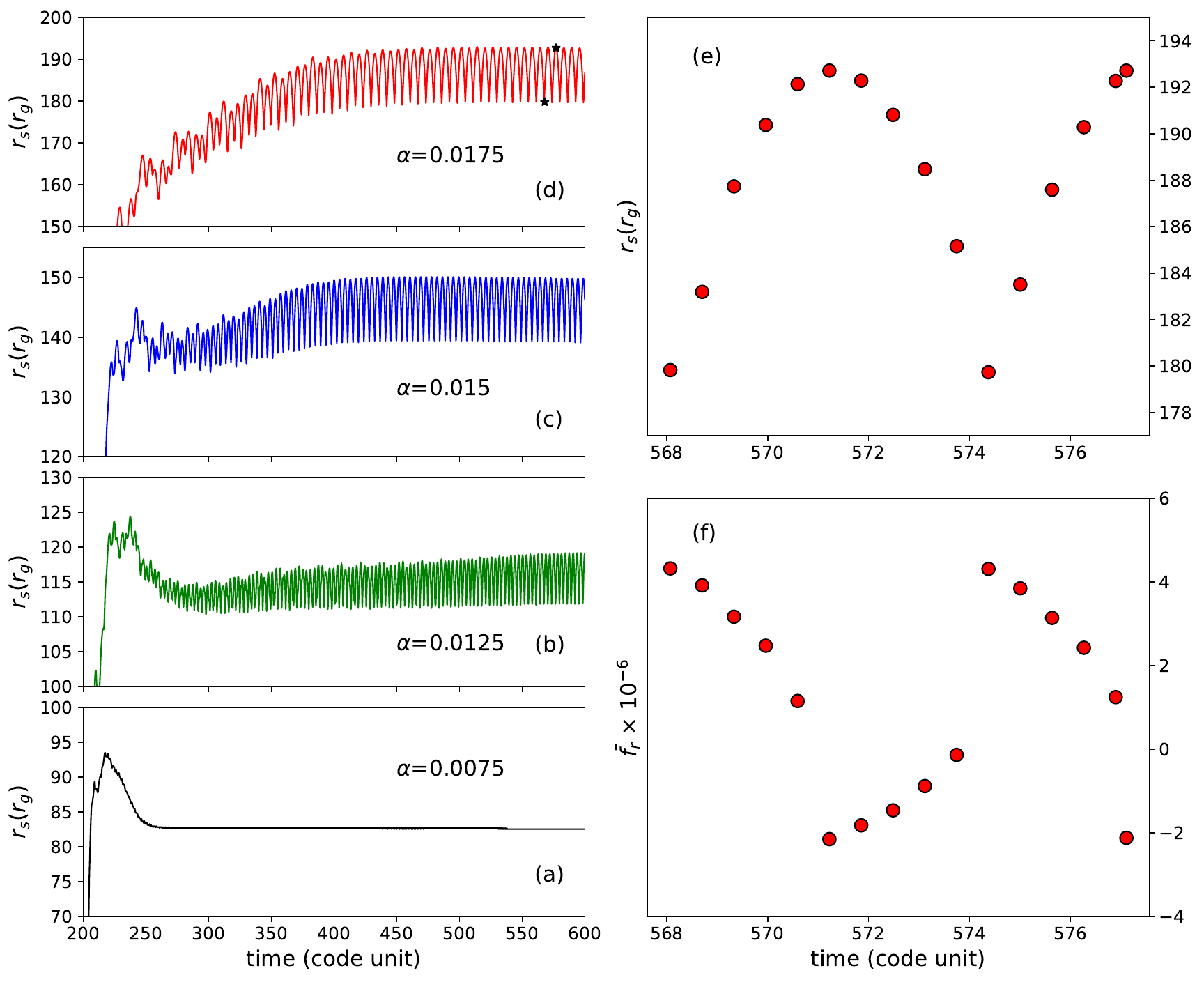}
        \caption{{{Time evolution of shock position for model E1. Viscosity parameters are (a) $\alpha=0.0075$, (b) $\alpha=0.0125$, (c) $\alpha=0.015$ and (d) $\alpha=0.0175$. In panel (e) $\rsh$ vs $t$ time series of panel (d) is plotted within a time bounded by two blue star, and panel (f) represent corresponding variation of the average force.}}
        }
        \label{fig:e2}
        \end{center}
\end{figure*}
We study the effect that viscosity might have on a shocked accretion flow for a given outer boundary condition. We start with an inviscid solution and then go on increasing $\alpha$. In the previous sections,
we injected with a relatively shorter boundary, although the injection was subsonic i. e., $M(\rou)<1$. In order to study shock oscillation, we would need a larger computation region. 
It is clear that the accretion solution crucially depends on
$\epsilon$ and $\lambda$. In {Fig. \ref{fig:same_inner}} we fixed the inner boundary ($\epsilon ~\&~\lambda_0$) and changed $\alpha$. Now we study how $\alpha$ affects accretion solutions with the same outer boundary. 

\subsection{Simulation set up}
\label{sec:setup}
Viscosity transports angular momentum outward, but importantly it is dependent on $\Theta$, apart from the shear. Therefore, it would be less effective in the supersonic region and more effective in the sub-sonic region. As a result, we see stronger angular momentum transport in the sub-sonic region. This implies there would be an accumulation of $\lambda$ in the post-shock region. 
Similarly, the dissipation of viscous heat in the post-shock disc is also greater compared to the pre-shock disc. In the viscous transonic flow, a shock is stationary if the total pressure (including the ram pressure, gas pressure, and rotational barrier) remains conserved across the shock front. As we increase the viscosity parameter $\alpha$, the angular momentum will be redistributed,
and therefore, beyond a certain critical value, it might destabilize the shock.

To properly study the effect of viscosity on accretion flow,
we considered six models L1, L2, L3 and E1, E2, E3 (see, Table \ref{tab:injection}). We obtain analytical inviscid solutions corresponding to $\epsilon=1.0001$, but for $\lambda=1.76,~, 1.77,~ 1.78$, and called them models L1, L2 and L3, respectively. From the analytical solutions, we chose the $\rou$, $\vou$, $\theou$, and $\lou$. Keeping these values fixed as the outer boundary, we then increased the viscosity parameter
$\alpha$ in the simulation code. The injection parameters of models E1, E2, and E3 on the other hand, is obtained for $\lambda=1.78$ and
$\epsilon=1.00005,~1.0001,~\&~1.00015$, respectively. From these analytical solutions the injection parameters $\rou$, $\vou$, $\theou$, and $\lou$ are chosen (see Table \ref{tab:injection}). Again, keeping these injection parameters fixed, we change the viscosity parameter $\alpha$. For all the models, we used 6826 uniform grids. The unit of time for models L1 --- L3 and E1---E3 is $t_{\rm unit}=10^{-2}(\mbh/M_{\odot})$s. 

All these models are chosen such that the respective inviscid solutions harbours shock. For all the models, we start with the injection
parameters with $\alpha=0$ till it reaches the steady state, and then we increase the $\alpha$ and study the changes induced by viscosity.
{Since all the models start with the inviscid flow, the angular momentum of the inviscid flow becomes the angular momentum at the outer boundary when viscosity is turned on. First, we study three cases where we kept $\epsilon$ same and studied flows with three different angular momentum, which we called L1 --- L3 models. Then we kept the angular momentum the same but studied cases for three different $\epsilon$ and called them models E1 --- E3.} 

\subsection{Models L1 --- L3}
For the model L1, 
once the steady state is reached in the inviscid limit, we found a standing stable shock at around $40.78r_g$. Then the viscosity is turned on in the simulation. Viscosity is more effective in the subsonic, hot region of the flow. And post-shock disc, being hot and subsonic, will transport $\lambda$ more efficiently, and there will be a pile-up of angular momentum in the post-shock disc.

This enhanced rotation of the disc near the shock front is likely to push the shock front outward.
If the inviscid flow has a steady shock, then a small amount of viscosity will eventually produce a steady shock in another position.
If the viscosity is increased further due to the piled up post-shock angular momentum, the shock position will overshoot the equilibrium position, and the shock starts to oscillate.
The critical viscosity for the model L1 for the shock starts to oscillate is {$\alpha=0.025$.}
Figure \ref{fig:variables} a-f, is the comparison of the flow variables $v_r$, $M$, {$\rho$, $p$, $\Theta$} and $\lambda$, respectively, from model L1.
The inviscid steady state solution (black, solid) is at code time $t_1=200$ and the time-dependent viscous flow for {$\alpha=0.035$} (red, solid)
was obtained at {$t_2=380$.} For the inviscid flow, the angular momentum is constant. However, for viscous flow, $\lambda$ is much higher than the pre-shock flow, which causes the shock oscillation.  
The $\lambda$ transport is much stronger in the post-shock region but in a small region near the shock front $d\lambda/dr \sim 0$. This also causes a small spike in $v_r$ and $M$, while a tiny dip in $\rho$ and $\Theta$ is also seen.
 
In Fig. \ref{fig:different_alpha}, we have shown the time evolution of the shock position for different $\alpha$ for model L2.  Figure \ref{fig:different_alpha}a shows the shock position with time for {$\alpha=0.01$}, {for which}
 a stationary shock forms at around {$61.02~r_g$.} {As viscosity is increased further}, the stationary shock becomes unstable, and the shock front starts to oscillate.
 Time evolution of shock position with the same outer boundary condition but different viscosity parameters {$\alpha=0.02$, $\alpha=0.025$ and $\alpha=0.03$} are shown in Fig. \ref{fig:different_alpha}b, \ref{fig:different_alpha}c and  \ref{fig:different_alpha}d, respectively.
 {In this model, there is significant shock oscillation for} $\alpha=0.03$. The shock oscillates with a mean position of about {$113.49r_g$ and has an amplitude of about $7.98r_g$.}
 
 \begin{figure*}
       \includegraphics[width=6 in]{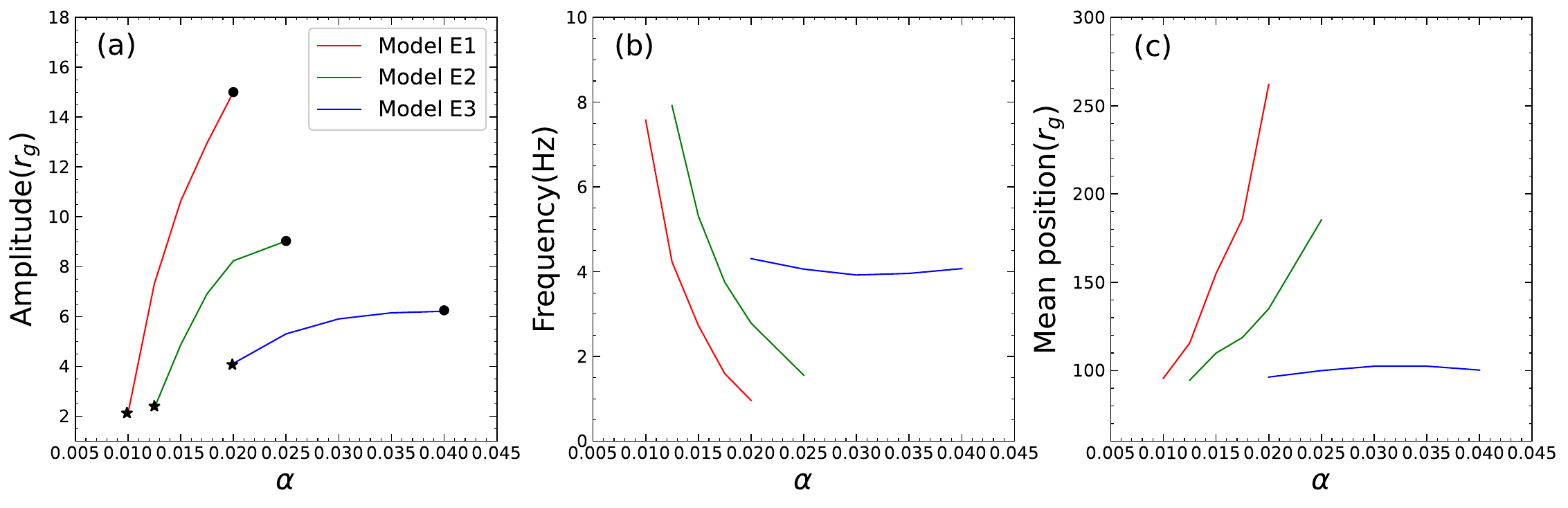}
        \caption{{{Variation of amplitude (a) and frequency of the shock oscillation (b) with $\alpha$ assuming black hole mass to be $M_{BH}=10M_{\odot}$. Flow parameter $\lou=1.78$. Red line is for model E1, blue line is for model E2, green line is for model E3.}}}
        \label{fig:alfa_var1}
\end{figure*}

\begin{figure*}
       \includegraphics[width=6 in]{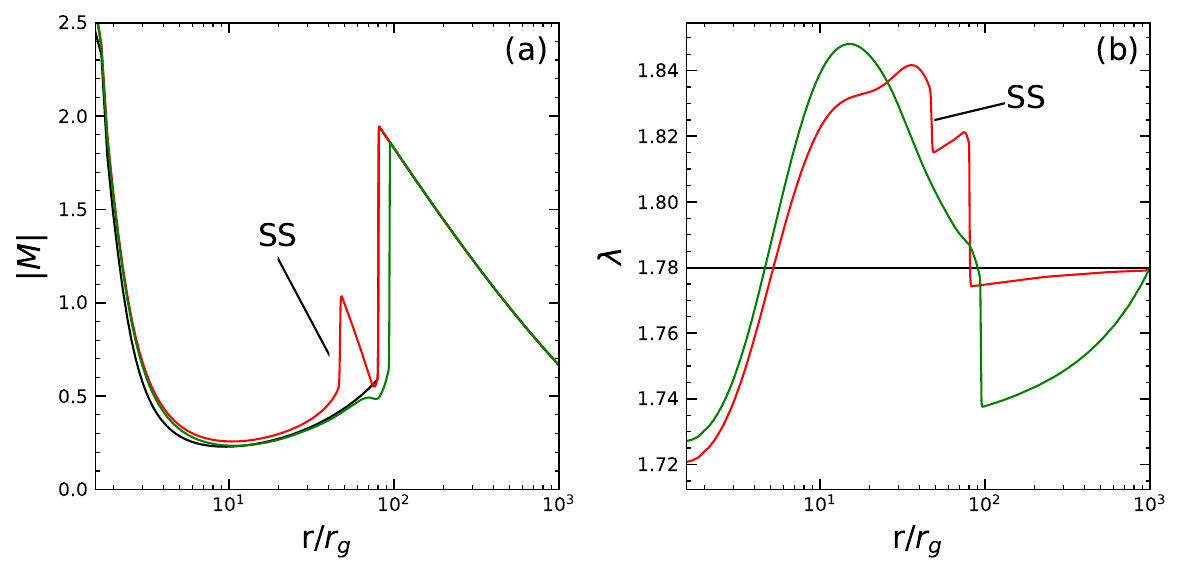}
        \caption{{{Snapshot of $M$ and $\lambda$ at times $t_1=348$ (black), $t_2=351$ (red), and $t_3=588$ (green) in code unit shown in (a) and (b). Time $t_1=348$ presents the inviscid steady state, $t_2=351$ present a dynamical state when a secondary shock (SS) forms, and in $t_3=588$ SS vanishes but primary shock oscillates for model E3. Viscosity parameter is $\alpha=0.02$.}}}
        \label{fig:multi_shock}
\end{figure*}

For model L3, the amplitude and the mean shock position increase while frequency decreases with increasing $\alpha$. In Fig. \ref{fig:alfa_var}a, we {compared} the amplitude of the shock oscillation as a function of $\alpha$ for models L1 (green), L2 (red), and L3 (blue).
{Comparison of the corresponding oscillation frequency is presented in} Fig. \ref{fig:alfa_var}b {and the mean position of oscillating shock is presented} in Fig. \ref{fig:alfa_var}c. The amplitude of the oscillation initially increases but tends to taper off with the increase of $\alpha$. The amplitude for a given $\alpha$ increases with increasing $\lou$. Frequency has an opposite trend, while the mean shock position follows the trend of the amplitude. {In Fig. \ref{fig:alfa_var}a, the oscillation starts at some critical viscosity parameter, say at $\alpha_l$ marked by the black star, while ends with a black dot, representing an upper limit of viscosity $\alpha_u$ beyond which there is no persistent oscillation. For model L2, $\alpha_l=0.016$ and $\alpha_u=0.038$. Therefore, the $\rsh$ vs $t$ plot for $\alpha=0.01$ in Fig. \ref{fig:different_alpha}a, shows no oscillation of the shock. If $\alpha> \alpha_u$, then the shock oscillates but propagates outward and eventually leaves the computation domain, indicating that the force balance do not allow for a persistent oscillation.}
\subsection{Models E1 --- E3}
In the case of models E1-E3, 
the injection parameters are given in Table \ref{tab:injection}, and the computational domain has a 6827 uniform grid. Each grid has a size of  $0.146r_g$, the same as model L1-L3. In other words, the resolution is kept exactly the same.
The injected angular momentum of models E1-E3 is exactly the same as that of model L3, but from E1 to E3, we inject progressively hotter and slower matter.
In Fig. \ref{fig:e2}a, \ref{fig:e2}b, \ref{fig:e2}c and  \ref{fig:e2}d, we plot the shock location $\rsh$ with time of model E1, for {$\alpha=0.0075$, $\alpha=0.0125$, $\alpha=0.015$ and $\alpha=0.0175$,} respectively. The critical viscosity which triggers shock oscillation is {$\alpha_l=0.0125$.} The
oscillation amplitude and mean shock position increased with increasing $\alpha$, but the frequency decreased. {We zoomed the $\rsh$ vs $t$ plot for $\alpha=0.0175$ (Fig. \ref{fig:e2}d) in Fig. \ref{fig:e2}e. We then compute the corresponding average acceleration ${\bar f}_r$ in the immediate post-shock flow at each time step and plotted them as a function of time. The average force is defined as 
$$
{\bar f}_r=\frac{\int^{r_*}_{\rsh} f_r dr}{\int^{r_*}_{\rsh} dr}
$$}
{Here $r_*$ is the position where the Mach number has its lowest magnitude or where $d|M|/dr \sim 0$. And $f_r=d\Phi/dr+\lambda^2/r^3-dp/(\rho dr)$ is the net radial acceleration. The acceleration is averaged from the shock location to the region
where the Mach number has the minimum value.
The ${\bar f}_r$ oscillates between positive and negative values, and it has a maximum value at the minimum shock location and a minimum value at the maximum $\rsh$ ({Fig. \ref{fig:e2}f}). And it is this oscillating average acceleration in the immediate post-shock region that causes the shock oscillation. The excess $\lambda$ pile up in the post-shock disc due to the presence of viscosity is driving the shock front outward. While as the post shock region expands the thermal gradient force becomes weaker, and gravity manages to restore the shock to its original location, which the $\rsh$ overshoots and oscillation continues. However, beyond a certain value of $\alpha$ i. e., $\alpha_u$, the median shock location continues to move outward, and the shock becomes unstable.}

In Fig. \ref{fig:alfa_var1}a, we plot and compare the amplitude of oscillation as a function of $\alpha$ for model E1 (red), model E2 (blue), and E3 (green). {The black stars and dots indicate the two critical viscosity parameters $\alpha_l$ and $\alpha_u$}. In Fig. \ref{fig:alfa_var1}b, we plot the corresponding frequencies, and in Fig. \ref{fig:alfa_var1}c, we plot the variation of mean position with $\alpha$. {The character of the oscillations in models} E1 and E2 are similar to the L3 model (i.e., amplitude/frequency increase/decrease with $\alpha$). For model E3 (green), the amplitude starts to increase with $\alpha$
but eventually saturates. The frequency, on the other hand, increases with $\alpha$, and it {saturates too}. The mean shock position
monotonically increases for E1 and E2, but
{remains roughly the same} for E3 with the increase of $\alpha$.
In Fig. \ref{fig:multi_shock}a \& b, we compare 
snapshots of $M$ and $\lambda$ at times {$t_1=348$ (black), $t_2=351$ (red) and $t_3=588$ (green) in code unit, for model E3 with $\alpha=0.02$.} Time $t_1=348$ represents the steady state, inviscid flow. At time $t_2=351$ (red), a secondary shock is observed to form. At time $t_3=588$ the secondary shock vanishes, but the primary shock continues to oscillate. It may be noted that the secondary shock is formed when the oscillating shock front ($\rsh$) approaches or exceeds $100 \rg$. When the shock front recedes to a large distance from the horizon, then the temperature distribution of the post-shock disc varies significantly. Therefore, closer to the horizon, the rate at which the angular momentum will be transported and the rate at which $\lambda$ would be transported near the shock front will be different and will cause one or two peaks in $\lambda$.
This causes formation of secondary shocks \citep[see,][]{2011ApJ...728..142L,2016ApJ...831...33L}. The tendency to form a secondary shock was also seen in Fig. \ref{fig:variables}, where a local dip in $\lambda$
corresponded with a local spike in $v_r$ and $M$, although the effect didn't develop into a secondary shock. In this figure, the secondary shock did form, but for a short time, and in later times, it washed away. This is quite different from the cases reported
in \cite{2011ApJ...728..142L}, where persistent secondary shock was reported. 

\begin{figure*}
	\begin{center}
       \includegraphics[width=6.0 in]{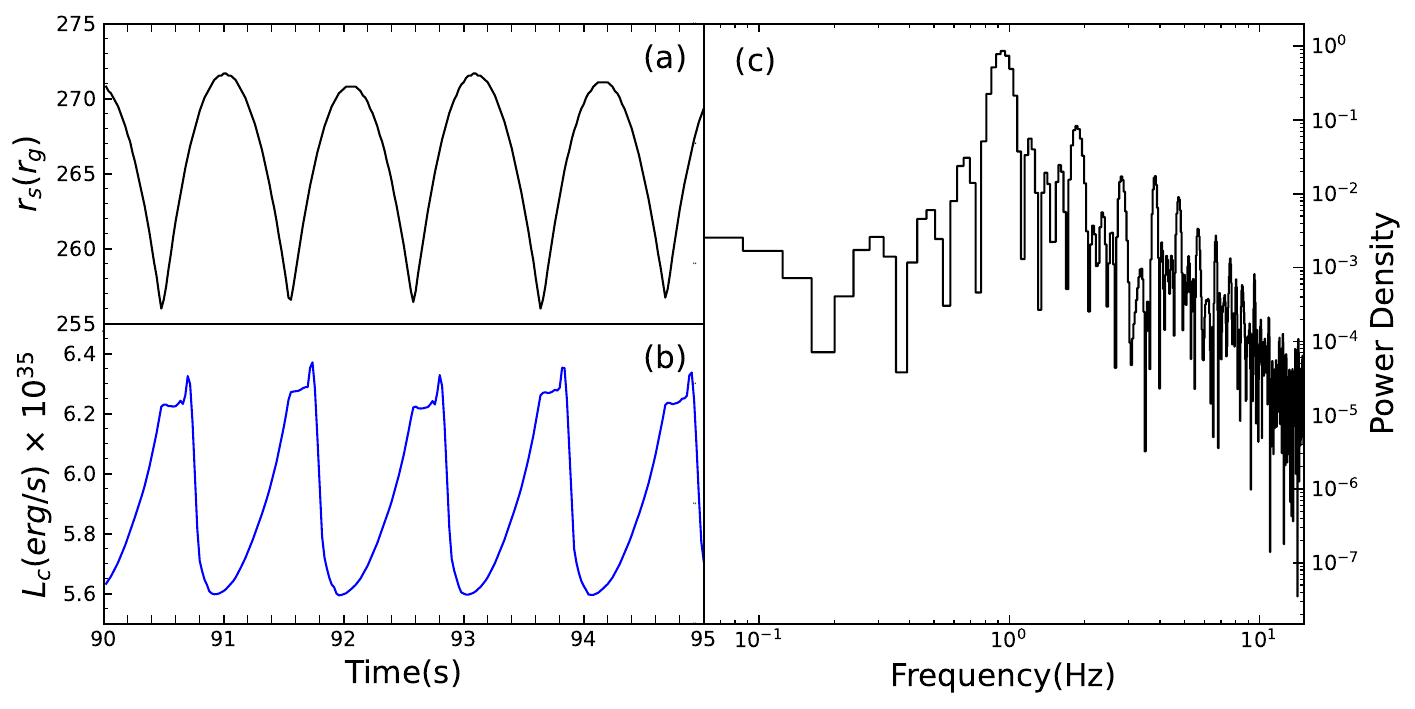}
        \caption{{{(a) Variation of shock position with time (in seconds), (b) Variation of total luminosity (${\mathrm{erg} s^{-1}}$) with time, (c) Power density spectrum (PDS)of the emission for the Model E1 and $\alpha=0.02$. The BH mass is $10~M_\odot$.}}}
        \label{fig:lofreqpds}
        \end{center}
\end{figure*}

\begin{figure*}
       \includegraphics[width=6 in]{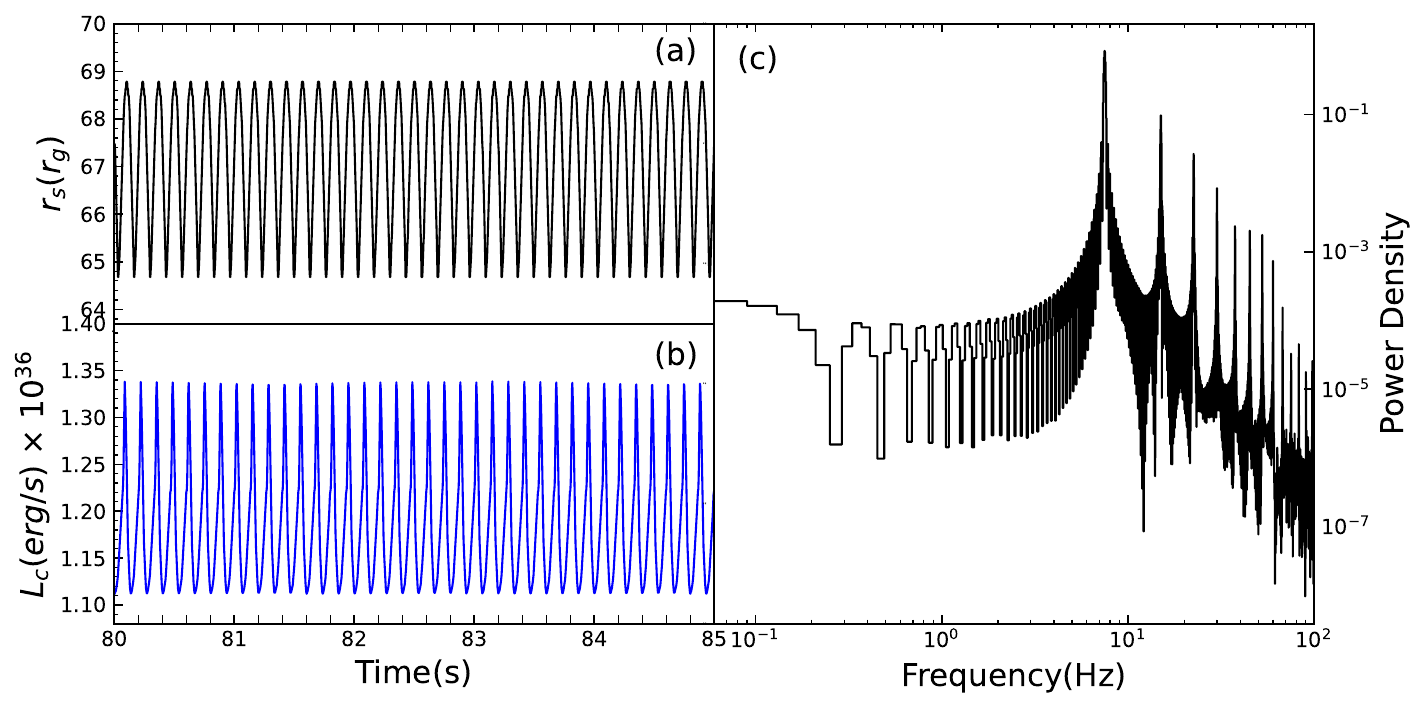}
        \caption{{{ Variation of shock position with time (in second) in (a), Variation of the total emission (${\mathrm{erg}~ \mathrm{s}^{-1}}$) with time in (b) for model L1 with $\alpha=0.03$. The power density spectrum of radiation variation is shown in (b) on logarithmic axes (c). Black hole mass is taken to be $M_{BH} = 10M_{\odot}$.}}}
        \label{fig:power}
\end{figure*}

\subsection{Effect of shock oscillation}
The post-shock is hotter and denser and so would emit high-energy photons. When the post-shock disc oscillates, the radiation emitted by it will also oscillate. Therefore, it is intriguing to investigate the time dependence of emission from such an oscillatory accretion disc. As radiative loss from the disc, we estimate the bremsstrahlung and synchrotron losses a posteriori from the disc as the representation of radiative processes. The emissivity resulting from bremsstrahlung (measured in ${\mathrm{erg}~\mathrm{cm}^{-3}~\mathrm{s}^{-1}}$) is described by \cite{1973blho.conf..343N} and
given as,
\begin{equation}
Q_{\rm br}=1.4\times10^{-27} n^2_e\sqrt{T_e}\left(1+4.4\times10^{-10}T_e\right)
 \label{eq:bremm}
\end{equation}
and the emissivity due to the synchrotron (in ${\mathrm{erg}~ {\rm cm}^{-3}~ {\rm s}^{-1}}$) is given by \cite{1983JBAA...93R.276S},
\begin{equation}
Q_{\rm syn}=\frac{16}{3} \frac{e^2}{c} \left( \frac{eB}{m_e c} \right)^2 \Theta^2_{e} n_e 
 \label{eq:synchro}
\end{equation}
At any time $t$, the total loss estimated posteriori would be
\begin{equation}
 L_c={\cal G}\int_{rsink}^{\rou}{\mathsf{F(\rsh)}}(Q_{\rm syn}+Q_{\rm br})r^2~dr 
 \label{eq:cool}
\end{equation}
Here, ${\mathsf{F(\rsh)}}$ represents the fitting function for the Comptonization parameter. This function is computed upon obtaining the accretion shock \citep[similar to][]{2014MNRAS.443.3444K}. The form of this analytic function is given by
\begin{equation}
{\mathsf{F(\rsh)}}=A\exp({\frac{-(\rsh-c_1)^2}{2c_2}})+B\exp({\frac{-\rsh}{c_3}})
 \label{eq:copmto}
\end{equation}
Where the values of parameters are $A=5.915, c_1=100.527, c_2= 59.314, c_3=-6131787530.355, B=1.972$.
For $Q_{\rm syn}$, the magnetic field is assumed stochastic and is estimated assuming the magnetic pressure is 1\% of the gas pressure.
We followed \cite{2014MNRAS.437.2992K} approach to estimate electron temperature. 
We assume the mass of the BH as
$\mbh=10M_\odot$. 

In Fig. \ref{fig:lofreqpds}, we plot the time series of shock oscillation (panel a), the time series of the total luminosity $L_c$
(panel b) and the corresponding power density spectrum (panel c), for the model E1 and {$\alpha=0.02$.} Assuming an accretion rate of $0.1 {\dot M}_{\rm Edd}$ we obtained the emission and then computed the power spectrum (PDS) for emission luminosity $L_c$ We obtained a fundamental frequency of {$\nu_{\rm fund}=0.964$ Hz and three harmonics.}
%

In Fig. \ref{fig:power} a and b, we plot the time series of shock position $\rsh$, the total power ($L_c$) for the model {L1 and $\alpha=0.03$.} The BH mass and accretion rate same as above.
Interestingly, there appears to be a phase gap between the 
oscillation in emission and shock oscillation.
To analyze the oscillation of radiation, we employ the methods described in \cite{2005A&A...431..391V}. The PDS of the radiation is depicted in Fig. \ref{fig:power}c. 
The PDS reveals that the quasi-periodic oscillation of radiation is characterized by the fundamental frequency {$\nu_{\rm fund} = 7.49$ Hz.} The first significant harmonics are observed at {15.07 Hz (approximately $2\nu_{\rm fund}$), 22.71 Hz (approximately $3\nu_{\rm fund}$), and 30.04 Hz (approximately $4\nu_{\rm fund}$) and so on.} 
{From Figs. \ref{fig:lofreqpds} and \ref{fig:power}, it is clear that the radiative efficiency of these flows is very low ($\sim 10^{-3}$). Realistic cooling may not qualitatively modify the results. We also included cooling in a very simplified manner and presented it in Appendix \ref{sec:cool}, which shows considering cooling may not dramatically change the solution}. 

It may be noted that the frequencies depicted in Figs. \ref{fig:alfa_var} \& \ref{fig:alfa_var1} are the fundamental frequencies of the shock oscillation. Although there is a phase shift between the oscillation of the radiation with that of the shock, the frequencies are the same. The peaks in emission are just after the shock minima. So, as the shock compresses the post-shock flow, making it hotter and denser the radiation peaks just after that. This has been seen before \cite{2011ApJ...728..142L,2016ApJ...831...33L,2014MNRAS.442..251D}.
Table \ref{tab:shock_properties} lists all the models and various parameters related to shock oscillation. 
The frequencies mentioned are the fundamentals of oscillation.
It shows that depending on the energy and angular momentum of the flow, the oscillations can range from frequencies that are less than Hz to a few Hz of frequencies.

\begin{table}
\centering
    \caption{Properties of shock oscillation for different models with different values of $\alpha$.}
    \label{tab:shock_properties}
    {\begin{tabular}{llccl}
    \hline
    Model & $\alpha$  & Amplitude  & Frequency & Mean Position\\
       &  & ($r_g$) & (Hz) &    ($r_g$)\\
    \hline
    \hline
    L1 & 0.03   &  4.089  & 7.485  & 66.733\\
    & 0.04  &  5.912    & 5.165 & 86.801 \\
    &0.045   &  6.953    & 3.891&  103.703\\
    &0.05   &  8.101  & 2.681 &  134.320\\
    \hline
    L2 & 0.02   &  4.558 & 5.570 & 80.125 \\
    & 0.025  & 6.397   & 4.369 &  95.320\\
    &0.03   &  7.975   & 3.230& 113.485	 \\
    &0.035   & 8.985 & 2.269 &  148.001\\
    \hline
    L3 & 0.015   &  4.864  & 5.314 & 109.894 \\
    & 0.0175 & 6.912 & 3.752 & 118.699 \\
    &0.02   &  8.235  & 2.786 & 135.018\\
    &0.025   &  9.032 & 1.557 & 185.337 \\
    \hline
    E1 & 0.0125   &  7.312  & 4.232 & 115.573\\
    & 0.015  &  10.630    & 2.732 &  154.945\\
    &0.0175   &  12.958  & 1.595 &  185.821\\
    &0.02   &  15.004  & 0.964 & 262.052 \\
    \hline
    E2 & 0.015   &  4.864  & 5.314 & 109.894 \\
    & 0.0175 & 6.912 & 3.752 & 118.699 \\
    &0.02   &  8.235  & 2.786 & 135.018\\
    &0.025   &  9.032 & 1.557 & 185.337 \\
    \hline
    E3 & 0.025   &  5.305 & 4.060 & 99.978	\\
    & 0.03 & 3.854 & 3.922	&  102.461\\
    &0.035   & 6.150&3.960 & 102.497\\
    &0.04   & 6.212& 4.073& 100.282\\
    \hline
    \end{tabular}}
\end{table}

\section{SUMMARY AND DISCUSSION}

In this study, we have conducted time-dependent numerical simulations to investigate the dynamics of viscous accretion flows surrounding black holes. 
For our investigations, we used a numerical code that implements the Eulerian TVD (Total Variation Diminishing) scheme along with a variable adiabatic index equation of state. Moreover, we have assumed a fully ionized electron-proton plasma. In addition, used the angular momentum density ($\rho~\lambda$) as one of the state variables and the specific angular momentum ($\lambda$) as one of the primitive variables and 
then solved the equations of motion. As a result, this code ensures strict conservation of angular momentum in the absence of viscosity and ensures proper angular momentum transfer
such that the angular momentum distribution is correct in the presence of viscosity. Therefore, with the use of CR EoS, we obtained a correct temperature distribution, and using the proper angular momentum equation instead of the equation for azimuthal speed, we obtained the correct angular momentum distribution within the framework of the Eulerian, upwind numerical scheme. Through rigorous testing, we demonstrated that the code accurately reproduces the angular momentum and temperature distribution similar to that in the analytical solution for viscous transonic flows (see Fig. \ref{fig:viscous_l}).  
In order to
be satisfied,
we regenerated all types of accretion solutions with our simulation code, and as solutions reached a steady state, we compared them with the analytical steady-state solutions for $\alpha \geq 0$. These were the tests of steady-state solution where $\alpha$ is fixed but for different $\epsilon$ \& $\lambda_0$.  
We also regenerated the analytical accretion solutions for the same inner boundary condition by changing the outer boundary condition for a given $\alpha$ with our simulation code. And then finally, we compared time-dependent accretion solutions by keeping the outer boundary the same but changing the $\alpha$.
We showed the accretions starting with no shock in the inviscid solution may harbour a shock due to the transfer of angular momentum. Of course, above critical viscosity, the shock will also disappear.
{The shock considered here is thin. In accretion discs, the $r-\phi$ component of the stress dominates over other components and that makes the shock front (if present) to be thin. In analytical/semi-analytical solutions, the shock front is an infinitesimally thin surface, but the numerical simulation code takes 5-6 cells to resolve the shock. Different dissipation processes may make the shock front spread over a small but finite width.}

In our time-dependent study, we focused specifically on the behaviour of shock in the presence of viscosity. We {considered} six models
to study them, three models varied the injection angular momentum but injection temperature and velocity is computed by keeping $\epsilon$ fixed, and three models varied the injection velocity and temperature according to varying $\epsilon$ values but at a given angular momentum at the outer boundary.
In this paper, it has been demonstrated that for low values of viscosity parameter, the shock front in a disc tends to settle into a position, which might be different from the inviscid case.  
In the case of higher viscosity, the rate of angular momentum transfer increased, leading to a faster expansion of the shock front. As the shock front surpassed a potential equilibrium position, it initiated oscillations (Fig. \ref{fig:different_alpha}, Fig. \ref{fig:e2}). This particular value of $\alpha$ can be denoted as critical viscosity. It is important to note that there is no specific value of the critical viscosity parameter but rather dependent on the initial outer boundary conditions ($~\vou, ~\theou, ~\lou$). For instance, as the $\lou$ and $\theou$ of the accretion flow increase for a given $\vou$, the value of critical viscosity decreases, but for increasing $\theou$ for a given value of $\lou$, it tends to have a higher critical viscosity parameter. This implies that the threshold at which the shock front undergoes oscillation is influenced by the specific characteristics of the flow, such as angular momentum and energy. In our simulations, we observed that the rate of outward angular momentum transport in the subsonic region is higher than in the supersonic region, and therefore, near the sink, the $\lambda$ distribution tends to flatten out. The post-shock $\lambda$ distribution is lower for steady-state, while in the case of oscillating shock, it is higher. It is the higher angular momentum transport that drives the shock oscillation.
{It has been shown in this paper that the viscosity transports angular momentum in the post shock disc more efficiently, causing the angular momentum to pile up.
If the piling up angular momentum is such that along with the thermal gradient term, the average acceleration of the mass of post shock fluid pushes the shock front outward. As the shock front moves outward the thermal
gradient force weakens, allowing gravity to bring it back, and then the shock front oscillates, generating what is called oscillating shocks}. For model L3 i. e., higher $\lou$, the mean shock position and the oscillation amplitude increase with increasing $\alpha$. 
For models E1-E2, which imply hotter flow at the outer boundary, the mean shock position and its amplitude decrease with the viscosity parameter. But for E3 the mean shock position decreases and then becomes flat with increasing $\alpha$.
The power density spectra of the time series of the emission for models E1 and {L1} show generally a C-type QPO with a higher Q factor and $L_c$ is also low, but the QPO for E1 is broader
compared to {L1}. Interestingly, the mean position of shock of model E1 with $\alpha=0.02$ is around few $\times ~100 \rg$, while for {L1 ($\alpha=0.03$) it is $\sim 67 \rg$.} Oscillating a larger mass of fluid induces a degradation in the quality of
oscillation.

{In our simulations, we have neglected radiative cooling. However, we kept an option of including a simpler form of cooling by reducing the total viscous heat dissipation by a factor $0\leq f_c <1$.
Typically similar to what was done by \cite{1994ApJ...428L..13N}. But in the bulk of the paper, we considered $f_c=0$, i.e., no cooling, in order to study only the effect of viscosity. However, we showed in the appendix that
even if cooling is considered to be $10\%$ of the viscous heating, the solution is affected marginally. The median shock location would go inward marginally, and the oscillation frequency would increase marginally.
However, the whole approach to cooling in this paper is very simplistic.
Posteriori computation of bremsstrahlung and synchrotron emission from these accretion flows is also very low. However, radiative cooling is a tricky topic, and the complications involved are beyond the scope of the present paper
\citep[see for example][]{2019IJMPD..2850037S,2020A&A...642A.209S,2023MNRAS.522.3735S}. We would take up this study in all seriousness in the near future.} 

There are few examples of simulations of viscous accretion flow in the literature \citep{1998MNRAS.299..799L,2011ApJ...728..142L,2014MNRAS.442..251D,2016ApJ...831...33L}. These previous papers were all studied with the supersonic injections and fixed $\Gamma$ EoS. Since viscosity is not very effective in the supersonic region, in all of these simulations, angular momentum distribution is almost constant or has a very smooth gradient in the pre-shock region. Viscosity is effective in the subsonic region, so in the subsonic post-shock region the angular momentum transport is very effective, resulting in a notable pile-up of angular momentum in the post-shock disc. Such a significant imbalance in angular momentum across the shock front causes large amplitude oscillation. On the contrary, in the present work, we injected the matter with subsonic velocities and used variable $\Gamma$ EoS. Moreover, the injection parameters were chosen from analytical inviscid solutions, which harbours a steady shock. It is well known that an accretion shock solution will pass through the outer sonic point, then a shock, and then enter the black hole through the inner sonic point. The angular momentum transport between injection point $\rou$  and the outer sonic point is significant, and it is also similar in the post-shock region. So, the 
angular momentum transport in the post-shock and pre-shock disc is comparable. As a result, the oscillation amplitude is much less than in the previous simulations with supersonic injections. Hence, when the mean shock position is greater than $100\rg$, even then, persistent secondary shocks do not form as was reported by \cite{2011ApJ...728..142L,2016ApJ...831...33L}. We found a temporary secondary shock may form (model E3, $\alpha=0.02$), but it got washed away. It may also be noted that the effect of shock oscillation does not travel beyond the outer sonic point, and therefore, the shock oscillation does not affect the outer sonic point and the outer boundary. In other words, the supersonic region between the shock and the outer sonic
point acoustically disconnects the shock oscillation from affecting the outer boundary. Therefore, as the shock oscillates, it is primarily the radiation from the shocked region that oscillates, and the radiation from the pre-shock region remains
relatively unaffected by shock oscillation. 

This is probably the first numerical simulation of viscous transonic accretion flows onto black holes using a variable $\Gamma$ EoS. Moreover, the dependence of properties of time-dependent accretion shocks on flow parameters is represented in Figs. \ref{fig:alfa_var} \& \ref{fig:alfa_var1}, are also probably obtained for the first time. The frequency of the shock oscillation may decrease or increase, with the increase of viscosity parameter depending on the energy and angular momentum of the flow, and similarly, the mean position of the oscillating shock can increase or decrease with the increase of viscosity parameter. Finally, low-frequency C-type QPO around a stellar-mass black hole, ranging from frequencies less than a Hz to tens of Hz, can be easily explained by oscillating shock.

\section*{Acknowledgements}
The authors acknowledge the anonymous referee for fruitful suggestions. 

\section*{Data Availability}
The data underlying this article will be shared on reasonable request to the corresponding author.



\bibliographystyle{mnras}
\bibliography{biblio} 




\appendix
\label{sec:cool}
\section{Shock oscillation with cooling}
\begin{figure}
    \centering
    \includegraphics[width=3.3 in]{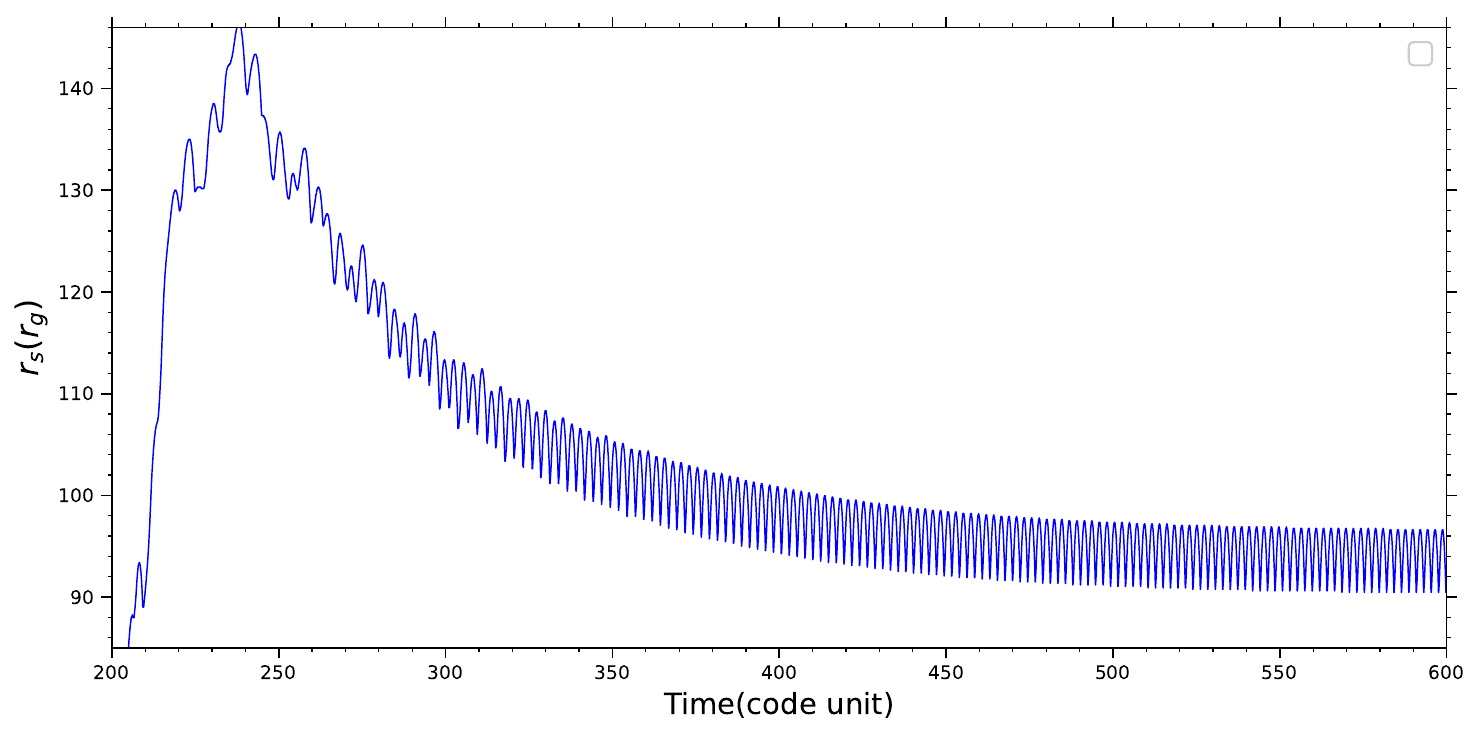}
    \caption{{Shock position with time for model L2 with $\alpha=0.025$ with cooling ($f_c$=0.1). Same model with same $\alpha$ without cooling shock oscillation is shown in Fig.\ref{fig:different_alpha}c.}}
    \label{fig:shock_cooling}
\end{figure}

{Here, we have studied the effect of viscosity on the oscillation of shocks in accretion flows using a 1D simulation code, considering the cooling effect. Cooling has been addressed similar to the way Narayan and his collaborators  {\citep{1994ApJ...428L..13N}} approached it, by incorporating a $(1-f_c)$ factor in the source term of the energy equation, where $f_c<1$ is the cooling factor. In our paper, we considered $f_c=0$. However, based on the posteriori calculation of bremsstrahlung and synchrotron emissions, it was evident that the radiative output was low (considering ${\dot M}=0.1{\dot M}{\rm Edd}$ and $M{\rm BH}=10 M_\odot$, with a radiative efficiency of $10^{-3}$). To explore the quantitative effect, we included $f_c=0.1$ and demonstrated its impact. The shock variation for model L2 with $\alpha-0.025$ and this level of cooling is shown in Fig. \ref{fig:shock_cooling}. Without considering cooling for this particular model, the shock oscillates with a frequency of 4.369$H_z$, an amplitude of 6.397$r_g$, and a mean position of 95.320$r_g$. However, with cooling, the shock oscillates with a frequency of 4.476$H_z$, an amplitude of 6.207$r_g$, and a mean position of 93.397$r_g$. Therefore, cooling would induce quantitative changes, but the conclusions would be qualitatively similar to what has been presented in the main part of the paper.}


\bsp	
\label{lastpage}
\end{document}